\documentclass[10pt,aps,prd,preprintnumbers,showpacs,superscriptaddress,nofootinbib,amsmath,amssymb,floats,floatfix,showkeys,notitlepage,longbibliography]{revtex4-2}
\usepackage{orcidlink}

\numberwithin{equation}{section}
\usepackage{comment}
\usepackage{booktabs}
\usepackage{stmaryrd}

\usepackage{threeparttable}
\usepackage{graphicx}
\usepackage{xcolor}
\usepackage{subfigure}
\usepackage{palatino}
\usepackage{sans}
\usepackage{hyperref}
\usepackage{etoolbox}
\hypersetup{colorlinks=true,linkcolor=blue,urlcolor=blue,citecolor=blue}
\usepackage[toc,page]{appendix}
\usepackage[normalem]{ulem}
\usepackage{adjustbox}
\usepackage{latexsym}
\usepackage{amsmath}
\usepackage{amssymb}
\usepackage{amsfonts}
\usepackage{dcolumn}
\usepackage{bm}
\usepackage{tikz}
\usepackage{bigints}
\usepackage{array,tabularx,multirow,booktabs}
\usepackage[tracking=true]{microtype}
\usepackage{soul} 
\SetTracking{}{500}
\SetTracking{encoding={*}, shape=sc}{40}
\UseRawInputEncoding 
\allowdisplaybreaks

\makeatletter
\patchcmd{\subequations}
  {\def\theequation{\theparentequation\alph{equation}}}
  {\def\theequation{\theparentequation\ifnum\c@equation>\z@.\alph{equation}\fi}}
  {}
  {\PackageError{main_final}{Unable to set dotted subequation numbering}{}}
\makeatother

\begin{document} \sloppy
\title{Energy, Passivity, and Diffusive Representations in Anisotropic Fractional Maxwell--Cole--Cole Media}

\author{Nikko John Leo S. Lobos \orcidlink{0000-0001-6976-8462}}
\email{nslobos@ust.edu.ph}
\affiliation{Electronics Engineering Department, University of Santo Tomas, Espa\~na Boulevard, Sampaloc, Manila 1008, Philippines}

\begin{abstract}
This study develops an energy formulation for linear anisotropic Maxwell--Cole--Cole media with multiple fractional polarization channels, passive conductivity, prescribed external currents, and finite-terminal Caputo memory with zero prehistory. A weighted temporal solution class admits the integrable fractional initial layer generated by quiescent polarization data. The fractional polarization power separates into a non-negative history-storage functional and a non-negative constitutive dissipation rate. The instantaneous dissipation may diverge at initialization but remains integrable, while the stored history energy and accumulated loss vanish initially. Combining this decomposition with the Poynting theorem gives a global field-plus-memory energy identity, electromagnetic passivity, closed-system energy decay, and conditional uniqueness under boundary conditions with zero supply. The analysis also recovers the Debye, instantaneous-dielectric, vanishing-susceptibility, and vacuum limits under stated convergence and compatibility assumptions. A positive diffusive representation gives a second admissible energy decomposition. Its storage does not exceed the history storage. Their difference is an exact non-negative weighted variance that vanishes only for a zero polarization history. The long-time limit of the finite-time harmonic response gives the anisotropic multichannel Cole--Cole permittivity, whose Hermitian imaginary part yields non-negative cycle-averaged absorption. A reproducible single-channel parameter study shows how the fractional order broadens the dielectric-loss band, redistributes wave attenuation, strengthens the integrable initial power layer, and lengthens the step-response tail. An exact electric-step response checks both energy representations. The resulting formulation relates the fractional initial layer, hereditary storage, passivity, diffusive relaxation, dispersion, absorption, and attenuation in a spatially local constitutive model.
\end{abstract}


\maketitle

\section{Introduction}
\label{sec:introduction}

Dielectric relaxation is hereditary when the present polarization depends on the past electric field. Cole and Cole introduced the fractional-power susceptibility as a phenomenological model of non-Debye dielectric dispersion \cite{ColeCole1941}. Jonscher described empirical patterns in broadband dielectric dispersion and absorption \cite{Jonscher1977}. Mainardi linked fractional kernels to Mittag--Leffler relaxation and distributed relaxation times \cite{Mainardi2010}. Garrappa, Mainardi, and Maione established completely monotone representations for major non-Debye relaxation laws, including the Cole--Cole model \cite{GarrappaMainardiMaione2016}. Li, Huang, and Lin developed a finite-element method for wave propagation in Cole--Cole media \cite{LiHuangLin2011}. Bai, Huang, Rui, and Wang constructed a fast second-order finite-difference time-domain method for time-fractional Maxwell systems \cite{BaiHuangRuiWang2022}. Yin, Liu, Li, and Zhang studied discrete energy dissipation for Maxwell--Cole--Cole equations \cite{YinLiuLiZhang2023}. Pietruszka, Stefa\'nski, and Gulgowski formulated a finite-difference time-domain scheme for time-fractional constitutive laws \cite{PietruszkaStefanskiGulgowski2023}. Baranowski, Bauer, and Mularczyk developed positive auxiliary-state realizations \cite{BaranowskiBauerMularczyk2021}. Diethelm \cite{Diethelm2022} and Chaudhary and Diethelm \cite{ChaudharyDiethelm2024} further studied convergence and computable diffusive representations. Jin, Quan, Wohlmuth, and Zhou analyzed the weak temporal regularity that often occurs near the initial time in fractional evolution problems \cite{JinQuanWohlmuthZhou2025}.

These studies provide constitutive models, time-domain solvers, auxiliary-state representations, and regularity theory. A separate finite-time question remains for initialized anisotropic Maxwell--Cole--Cole media. Polarization power must be divided into admissible storage and dissipation when the material has several fractional relaxation channels, memory begins at a finite time, and the Maxwell boundary supply is retained. This study formulates that energy problem, derives history and diffusive descriptions from the same constitutive law, and compares the two descriptions. The focus is the continuum energy structure, not a new fractional Maxwell equation or time-stepping algorithm.

For $0<\alpha<1$, a finite-terminal Caputo derivative weights the post-initialization rate of a state variable with the weakly singular kernel $t^{-\alpha}/\Gamma(1-\alpha)$ \cite{Caputo1967}. The lower terminal is part of the initial-value model. At finite time, a medium initialized at $t=0$ is not equivalent to a stationary fractional operator with an infinite prehistory. Lorenzo and Hartley formulated initialized fractional calculus and described the forcing produced by prehistory \cite{LorenzoHartley2000}. Hartley and Lorenzo later analyzed initialized fractional-order systems and the persistence of initial memory \cite{HartleyLorenzo2002}.

In this dielectric model, the Caputo operator acts only in the temporal polarization law. The Caputo derivative of a constant vanishes, so ordinary electromagnetic initial data are preserved while the polarization retains its post-initialization history. The fractional order $\alpha$ is a phenomenological relaxation parameter. It measures the departure from a single-time Debye response and the breadth of the effective relaxation-time distribution. The limit $\alpha\to1^{-}$ gives the Debye form. The order is not a fractional spatial dimension or a universal microscopic constant. Multichannel Cole--Cole fits to measured dielectric spectra, including biological-tissue parametrizations, show how the orders, strengths, and relaxation times enter material characterization \cite{GabrielLauGabriel1996}.

The model uses the standard macroscopic Maxwell equations and places the fractional response in the dielectric polarization. The medium is linear, nonmagnetic, spatially local, and temporally dispersive. It may contain several anisotropic Cole--Cole channels with distinct fractional orders and relaxation times. Passive conductivity and prescribed external current remain separate because they have different roles in the electromagnetic power balance. The constitutive tensors may be discontinuous across Lipschitz interfaces, with the Maxwell equations interpreted in the corresponding weak setting \cite{Monk2003}.

The unresolved issue is the finite-time meaning of stored and dissipated polarization energy, not the existence of a complex permittivity. In a hereditary material, this division depends on the history of the internal state and cannot be determined from a frequency-domain susceptibility alone. Several non-negative storage or free-energy functionals may be compatible with the same constitutive law \cite{Fabrizio:2014,DeseriDiPaolaZingales2014,AmendolaFabrizioGolden2021}. An energy formulation must therefore state its initialization, regularity class, supply convention, and physical scope.

The analysis first defines the anisotropic multichannel Maxwell--Cole--Cole initial-value problem and a temporal function class that admits the weak fractional initial layer. It then derives a history-based power decomposition and includes it in the Maxwell energy balance. The limiting constitutive cases are examined under stated convergence assumptions. A positive diffusive representation of the same Caputo kernel is constructed and compared with the history representation. Finite-time harmonic and electric-step problems then connect the energy formulation with dispersion, attenuation, and measurable constitutive responses.

The wave and step-response section includes a reproducible single-channel parameter study. The fractional order varies while the other constitutive parameters remain fixed. Exact constitutive expressions show how the order affects frequency-domain loss, propagation attenuation, transient polarization, and fractional power. The material parameters are illustrative rather than fitted data. The calculation is not a convergence study of a fully discrete Maxwell solver.

The scope is restricted to zero prehistory, linear polarization, time-independent passive material tensors, spatially local response, and nonmagnetic media. The model excludes nonlinear polarization, magnetic dispersion, spatial nonlocality, aging coefficients, initialized nonzero memory, and thermal transport. The constitutive loss is an electromagnetic energy-dissipation rate. A statement about thermodynamic entropy production would require a thermal balance and a Clausius--Duhem analysis.

Section~\ref{sec:fractional_model} defines the model, notation, and weak solution class. Section~\ref{sec:energy_decomposition} derives the history energy balance. Section~\ref{sec:global_energy} states the global theorem and examines limiting cases. Section~\ref{sec:diffusive_representation} introduces the diffusive representation and compares it with the history formulation. Section~\ref{sec:wave_validation} examines harmonic propagation, the fractional-order parameter study, and electric-step consistency checks. Section~\ref{sec:conclusion} summarizes the results.

\section{Fractional Maxwell--Cole--Cole model}
\label{sec:fractional_model}
Let $\Omega\subset\mathbb{R}^{3}$ be a bounded Lipschitz domain occupied by a linear dielectric medium, and let $t\in[0,T]$, where $T>0$ is the observation time. The electric field $\mathbf{E}(\mathbf{r},t)$, magnetic field $\mathbf{H}(\mathbf{r},t)$, electric displacement $\mathbf{D}(\mathbf{r},t)$, and magnetic flux density $\mathbf{B}(\mathbf{r},t)$ satisfy the standard macroscopic Maxwell equations in SI units \cite{Jackson:1999},
\begin{subequations}
\label{eq:maxwell_system}
\begin{align}
\nabla\times\mathbf{E}
&=
-\partial_t\mathbf{B},
&
\nabla\times\mathbf{H}
&=
\mathbf{J}_{f}+\partial_t\mathbf{D},
\label{eq:maxwell_curl}
\\
\nabla\cdot\mathbf{D}
&=
\rho_f,
&
\nabla\cdot\mathbf{B}
&=
0,
\label{eq:maxwell_divergence}
\end{align}
\end{subequations}
where $\mathbf{r}\in\Omega$ is position, $\partial_t$ denotes time differentiation, $\nabla\times$ and $\nabla\cdot$ denote curl and divergence, $\rho_f$ is the free-charge density, and $\mathbf{J}_{f}$ is the free-current density. Equations~\eqref{eq:maxwell_curl} and \eqref{eq:maxwell_divergence} give the curl evolution laws and divergence constraints of the macroscopic Maxwell system. Taking the divergence of the Amp\`ere--Maxwell equation and using Gauss's law gives
\begin{equation}
\partial_t\rho_f+\nabla\cdot\mathbf{J}_{f}=0,
\label{eq:charge_continuity}
\end{equation}
so the fractional constitutive law does not alter local free-charge conservation.
The medium is nonmagnetic, spatially local, and temporally dispersive. Using the standard macroscopic constitutive framework \cite{Jackson:1999}, the present model separates the material response as
\begin{subequations}
\label{eq:constitutive_relations}
\begin{align}
\mathbf{B}
&=
\mu_0\mathbf{H},
\label{eq:magnetic_constitutive_relation}
\\
\mathbf{D}
&=
\epsilon_0\boldsymbol{\epsilon}_{\infty}\mathbf{E}
+
\sum_{j=1}^{N}\mathbf{P}_{j},
\label{eq:electric_displacement_relation}
\\
\mathbf{J}_{f}
&=
\boldsymbol{\sigma}\mathbf{E}
+
\mathbf{J}_{\mathrm{ext}},
\label{eq:free_current_relation}
\end{align}
\end{subequations}
where $\mu_0$ and $\epsilon_0$ are the vacuum permeability and permittivity, $\boldsymbol{\epsilon}_{\infty}(\mathbf{r})$ is the dimensionless high-frequency relative-permittivity tensor, $\boldsymbol{\sigma}(\mathbf{r})$ is the conductivity tensor, and $\mathbf{J}_{\mathrm{ext}}(\mathbf{r},t)$ is an externally prescribed current. The integer $N\geq1$ is the number of relaxation channels. The index $j=1,\ldots,N$ labels a channel, and $\mathbf{P}_{j}(\mathbf{r},t)$ is its polarization density. Equation~\eqref{eq:constitutive_relations} separates the vacuum magnetic response, high-frequency electric displacement, relaxing polarization, Ohmic conduction, and imposed current. Different channels may have distinct susceptibility tensors, relaxation times, and fractional orders. The isotropic single-channel reduction gives the standard time-domain Cole--Cole model \cite{ColeCole1941,GarrappaMainardiMaione2016,LiHuangLin2011,BaiHuangRuiWang2022}.
The conductive power density $\mathbf{E}\cdot\boldsymbol{\sigma}\mathbf{E}$ is nonnegative for a passive medium. Under the sign convention in Eq.~\eqref{eq:maxwell_curl}, positive $\mathbf{E}\cdot\mathbf{J}_{\mathrm{ext}}$ removes energy from the electromagnetic field, whereas a negative value supplies energy to it.

For $0<\alpha<1$, the finite-terminal Caputo derivative of an absolutely continuous vector-valued function $\mathbf{f}$ is defined componentwise by \cite{Caputo1967,Mainardi2022}
\begin{equation}
{}^{C}_{0}D_t^{\alpha}\mathbf{f}(t)
=
\frac{1}{\Gamma(1-\alpha)}
\int_{0}^{t}
(t-s)^{-\alpha}
\partial_s\mathbf{f}(s)\,ds,
\label{eq:caputo_definition}
\end{equation}
where $\Gamma$ is the Euler gamma function and $s$ is the history variable. The lower terminal $t=0$ sets the initialization time. No polarization history or auxiliary memory state is assigned for $s<0$. A nonzero prehistory would add hereditary forcing to the constitutive equation and contribute an initial memory energy \cite{LorenzoHartley2000,HartleyLorenzo2002}.
Starting from the established scalar Cole--Cole evolution equation \cite{ColeCole1941,GarrappaMainardiMaione2016}, the present model assigns one such law to each anisotropic polarization channel:
\begin{equation}
\tau_j^{\alpha_j}
{}^{C}_{0}D_t^{\alpha_j}\mathbf{P}_{j}
+
\mathbf{P}_{j}
=
\epsilon_0\boldsymbol{\chi}_{j}\mathbf{E},
\qquad
j=1,\ldots,N,
\label{eq:cole_cole_constitutive}
\end{equation}
where $\tau_j>0$ is the characteristic relaxation time of channel $j$, $0<\alpha_j<1$ is its fractional order, and $\boldsymbol{\chi}_{j}(\mathbf{r})$ is its dimensionless susceptibility tensor. The left side of Eq.~\eqref{eq:cole_cole_constitutive} combines hereditary relaxation with the instantaneous polarization amplitude. The right side is the electric driving term. The factor $\tau_j^{\alpha_j}$ ensures dimensional consistency. For a scalar, single-channel, time-translation-invariant response, Eq.~\eqref{eq:cole_cole_constitutive} gives the Cole--Cole susceptibility and its Mittag--Leffler relaxation law. In the notation used here, $\alpha_j$ is the fractional exponent in $[1+(s\tau_j)^{\alpha_j}]^{-1}$. This convention differs from Cole--Cole parametrizations that denote the exponent by $1-\alpha_{\mathrm{CC}}$.
Equation~\eqref{eq:cole_cole_constitutive} is a temporal coarse-graining of unresolved relaxation processes. The memory kernel weights each post-initialization polarization rate, $\partial_s\mathbf{P}_j$, by the age-dependent factor $(t-s)^{-\alpha_j}$. The tensor $\boldsymbol{\chi}_j$ sets the channel strength and directional response, $\tau_j$ sets its characteristic time, and $\alpha_j$ controls the persistence and breadth of the memory. The fractional operator therefore belongs in the material law rather than in Maxwell's spatial curl equations.
For a constant applied electric field, the transient solution approaches $\mathbf{P}_{j}=\epsilon_0\boldsymbol{\chi}_{j}\mathbf{E}$. The relaxed polarization permittivity is therefore
\begin{equation}
\boldsymbol{\epsilon}_{s}
=
\boldsymbol{\epsilon}_{\infty}
+
\sum_{j=1}^{N}\boldsymbol{\chi}_{j}.
\label{eq:static_permittivity_tensor}
\end{equation}
When $\boldsymbol{\sigma}\neq\mathbf{0}$, this tensor describes the relaxed polarization after the conductive current has been separated. It is not the zero-frequency limit of the full effective complex permittivity, because the conductive contribution scales as $\omega^{-1}$.
The material tensors are real, time independent, symmetric, and essentially bounded. Here $L^{\infty}(\Omega)$ denotes the space of essentially bounded coefficient fields, $\mathbb{R}_{\mathrm{sym}}^{3\times3}$ denotes the real symmetric $3\times3$ matrices, and $\mathbf{a}\in\mathbb{R}^{3}$ is an arbitrary test vector. The positive constants $c_{\infty}$ and $c_j$ express uniform positive definiteness. For almost every $\mathbf{r}\in\Omega$, the tensors satisfy
\begin{subequations}
\label{eq:material_hypotheses}
\begin{align}
\boldsymbol{\epsilon}_{\infty},
\boldsymbol{\chi}_{j},
\boldsymbol{\sigma}
&\in
L^{\infty}
\bigl(
\Omega;\mathbb{R}_{\mathrm{sym}}^{3\times3}
\bigr),
\label{eq:bounded_symmetric_material_tensors}
\\
\mathbf{a}\cdot
\boldsymbol{\epsilon}_{\infty}(\mathbf{r})\mathbf{a}
&\geq
c_{\infty}|\mathbf{a}|^{2},
\label{eq:positive_high_frequency_permittivity}
\\
\mathbf{a}\cdot
\boldsymbol{\chi}_{j}(\mathbf{r})\mathbf{a}
&\geq
c_j|\mathbf{a}|^{2},
\label{eq:positive_susceptibility}
\\
\mathbf{a}\cdot
\boldsymbol{\sigma}(\mathbf{r})\mathbf{a}
&\geq
0,
\label{eq:passive_conductivity}
\\
c_{\infty}
&>0,
\qquad
c_j>0.
\label{eq:positive_material_bounds}
\end{align}
\end{subequations}
Thus, $\boldsymbol{\epsilon}_{\infty}$ and $\boldsymbol{\chi}_{j}$ are uniformly positive definite, while $\boldsymbol{\sigma}$ is positive semidefinite. In particular,
$\boldsymbol{\chi}_{j}^{-1}\in
L^{\infty}(\Omega;\mathbb{R}_{\mathrm{sym}}^{3\times3})$,
and the susceptibility-weighted norm is
$|\mathbf{a}|_{\boldsymbol{\chi}_{j}^{-1}}^{2}
=
\mathbf{a}\cdot\boldsymbol{\chi}_{j}^{-1}\mathbf{a}$.
A semidefinite susceptibility may instead be treated on its active subspace. Uniform positive definiteness is imposed because the polarization energy contains $\boldsymbol{\chi}_{j}^{-1}$.
No spatial differentiability of the material tensors is required. The coefficients may be piecewise discontinuous across finitely many Lipschitz interfaces. Maxwell's equations are then understood almost everywhere within each material subdomain and distributionally on $\Omega$. In the absence of prescribed free surface charge and surface current, the weak formulation gives the standard transmission conditions \cite{Monk2003}
\begin{subequations}
\label{eq:interface_conditions}
\begin{align}
\llbracket\mathbf{n}\times\mathbf{E}\rrbracket
&=
\mathbf{0},
\label{eq:tangential_electric_continuity}
\\
\llbracket\mathbf{n}\times\mathbf{H}\rrbracket
&=
\mathbf{0},
\label{eq:tangential_magnetic_continuity}
\\
\llbracket\mathbf{n}\cdot\mathbf{D}\rrbracket
&=
0,
\label{eq:normal_displacement_continuity}
\\
\llbracket\mathbf{n}\cdot\mathbf{B}\rrbracket
&=
0,
\label{eq:normal_magnetic_flux_continuity}
\end{align}
\end{subequations}
where $\mathbf{n}$ is the interface normal and
$\llbracket \cdot \rrbracket$ denotes the jump. Surface sources, when present, must be included in the distributional charge and current densities.
The initial polarization is quiescent, while the electromagnetic fields are prescribed by
\begin{subequations}
\label{eq:initial_data}
\begin{align}
\mathbf{P}_{j}(\mathbf{r},0)
&=
\mathbf{0},
\label{eq:initial_polarization}
\\
\mathbf{E}(\mathbf{r},0)
&=
\mathbf{E}_{0}(\mathbf{r}),
\label{eq:initial_electric_field}
\\
\mathbf{H}(\mathbf{r},0)
&=
\mathbf{H}_{0}(\mathbf{r}),
\label{eq:initial_magnetic_field}
\end{align}
\end{subequations}
with the divergence constraints in Eq.~\eqref{eq:maxwell_divergence} imposed distributionally at $t=0$.
Quiescent memory does not imply that $\partial_t\mathbf{P}_{j}$ is bounded at initialization. For $0<\alpha<1$, define
\begin{equation}
\begin{split}
\mathcal{X}_{\alpha}
\bigl(0,T;L^{2}(\Omega)^{3}\bigr)
=
\Bigl\{
\mathbf{v} :
&
\mathbf{v}\in
C\bigl([0,T];L^{2}(\Omega)^{3}\bigr)
\cap
C^{1}\bigl((0,T];L^{2}(\Omega)^{3}\bigr),
\\
&
\mathbf{v}(0)=\mathbf{0},
\qquad
\sup_{0<t\leq T}
t^{1-\alpha}
|\partial_t\mathbf{v}(t)|_{L^{2}(\Omega)}
<\infty
\Bigr\}.
\end{split}
\label{eq:weighted_fractional_class}
\end{equation}
Assume that
$\mathbf{P}_{j}\in
\mathcal{X}_{\alpha_j}(0,T;L^{2}(\Omega)^{3})$.
In Eq.~\eqref{eq:weighted_fractional_class}, $C$ denotes time-continuous functions, $C^{1}$ denotes continuously differentiable functions on $(0,T]$, and $L^{2}(\Omega)^{3}$ denotes square-integrable vector fields on $\Omega$. The weighted supremum controls the allowed initial singularity of the polarization rate. This space permits
$\partial_t\mathbf{P}_{j}(t)=
\mathcal{O}(t^{\alpha_j-1})$
as $t\to0^{+}$. Since $\alpha_j>0$, the singularity is integrable and
$\mathbf{P}_{j}\in
W^{1,1}(0,T;L^{2}(\Omega)^{3})$,
so the Caputo convolution is well defined for almost every $t>0$. Weak initial singularities of this type are characteristic of Caputo evolution equations even for regular data \cite{JinQuanWohlmuthZhou2025}.
The initial layer follows directly from Eq.~\eqref{eq:cole_cole_constitutive}. The standard Riemann--Liouville fractional integral is \cite{Mainardi2022}
\begin{equation}
(I^{\alpha}\mathbf{f})(t)
=
\frac{1}{\Gamma(\alpha)}
\int_{0}^{t}(t-s)^{\alpha-1}\mathbf{f}(s)\,ds
\end{equation}
Using the composition identity $I^{\alpha}\,{}^{C}_{0}D_t^{\alpha}\mathbf{f}=\mathbf{f}-\mathbf{f}(0)$ \cite{Mainardi2022} and $\mathbf{P}_{j}(0)=\mathbf{0}$ gives the equivalent Volterra equation
\begin{equation}
\mathbf{P}_{j}
+
\tau_j^{-\alpha_j}
I^{\alpha_j}\mathbf{P}_{j}
=
\epsilon_0\tau_j^{-\alpha_j}
I^{\alpha_j}
\bigl(\boldsymbol{\chi}_{j}\mathbf{E}\bigr).
\end{equation}
If
$\mathbf{E}(t)=\mathbf{E}_{0}+o_{L^{2}}(1)$
as $t\to0^{+}$, and $\mathbf{E}$ is H\"older continuous when the differentiated expansion is used, then
\begin{subequations}
\label{eq:polarization_initial_layer}
\begin{align}
\mathbf{P}_{j}(t)
&=
\frac{
\epsilon_0\boldsymbol{\chi}_{j}\mathbf{E}_{0}}
{\tau_j^{\alpha_j}\Gamma(1+\alpha_j)}
t^{\alpha_j}
+
o_{L^{2}}(t^{\alpha_j}),
\label{eq:polarization_initial_asymptotic}
\\
\partial_t\mathbf{P}_{j}(t)
&=
\frac{
\epsilon_0\boldsymbol{\chi}_{j}\mathbf{E}_{0}}
{\tau_j^{\alpha_j}\Gamma(\alpha_j)}
t^{\alpha_j-1}
+
o_{L^{2}}(t^{\alpha_j-1}).
\label{eq:polarization_rate_initial_asymptotic}
\end{align}
\end{subequations}
A nonzero initial electric field therefore produces an integrable fractional layer in general. The polarization remains continuous at $t=0$ but is not generally continuously differentiable there.
The restrictive conditions required for classical temporal regularity follow from the same constitutive law. If
$\mathbf{P}_{j}\in
C^{1}([0,T];L^{2}(\Omega)^{3})$,
then
${}^{C}_{0}D_t^{\alpha_j}\mathbf{P}_{j}(t)\to\mathbf{0}$
as $t\to0^{+}$. Equation~\eqref{eq:cole_cole_constitutive} and the quiescent data then imply
$\boldsymbol{\chi}_{j}\mathbf{E}_{0}=\mathbf{0}$ and hence
$\mathbf{E}_{0}=\mathbf{0}$.
If, in addition,
$\mathbf{P}_{j}\in C^{2}([0,T];L^{2}(\Omega)^{3})$
and
$\mathbf{E}\in C^{1}([0,T];L^{2}(\Omega)^{3})$,
then the standard differentiation formula for the Caputo derivative \cite{Mainardi2022} gives, for $t>0$,
\begin{equation}
\partial_t
\bigl(
{}^{C}_{0}D_t^{\alpha_j}\mathbf{P}_{j}
\bigr)
=
\frac{
t^{-\alpha_j}\partial_t\mathbf{P}_{j}(0)}
{\Gamma(1-\alpha_j)}
+
\frac{1}{\Gamma(1-\alpha_j)}
\int_{0}^{t}
(t-s)^{-\alpha_j}
\partial_s^{2}\mathbf{P}_{j}(s)\,ds.
\end{equation}
The differentiated constitutive equation has a finite right-hand side at $t=0$. The singular first term therefore requires
$\partial_t\mathbf{P}_{j}(0)=\mathbf{0}$.
Taking $t\to0^{+}$ in the remaining equation and using the positive definiteness of $\boldsymbol{\chi}_{j}$ gives
\begin{subequations}
\label{eq:classical_compatibility_conditions}
\begin{align}
\boldsymbol{\chi}_{j}\mathbf{E}_{0}
&=
\mathbf{0},
\label{eq:classical_electric_compatibility}
\\
\partial_t\mathbf{P}_{j}(0)
&=
\mathbf{0},
\label{eq:classical_polarization_rate_compatibility}
\\
\partial_t\mathbf{E}(0)
&=
\mathbf{0}.
\label{eq:classical_electric_rate_compatibility}
\end{align}
\end{subequations}
These classical compatibility conditions are not assumed. The model retains the weak initial layer in Eq.~\eqref{eq:polarization_initial_layer}.
For the weak electromagnetic formulation, define
\begin{equation}
H(\operatorname{curl};\Omega)
=
\left\{
\mathbf{v}\in L^{2}(\Omega)^{3}:
\nabla\times\mathbf{v}\in L^{2}(\Omega)^{3}
\right\},
\end{equation}
and define $H(\operatorname{div};\Omega)$ by replacing the curl with the divergence. The fields and source are assumed to satisfy
\begin{subequations}
\label{eq:weak_field_regularities}
\begin{align}
\mathbf{E},\mathbf{H}
&\in
C\bigl([0,T];L^{2}(\Omega)^{3}\bigr)
\cap
L^{\infty}
\bigl(0,T;H(\operatorname{curl};\Omega)\bigr),
\label{eq:weak_electromagnetic_field_regularities}
\\
\partial_t\mathbf{E},
\partial_t\mathbf{H}
&\in
L^{1}\bigl(0,T;L^{2}(\Omega)^{3}\bigr),
\label{eq:weak_electromagnetic_rate_regularities}
\\
\mathbf{D},\mathbf{B}
&\in
L^{\infty}
\bigl(0,T;H(\operatorname{div};\Omega)\bigr),
\label{eq:weak_flux_regularities}
\\
\mathbf{J}_{\mathrm{ext}}
&\in
L^{1}\bigl(0,T;L^{2}(\Omega)^{3}\bigr).
\label{eq:weak_external_current_regularity}
\end{align}
\end{subequations}
These assumptions admit the $t^{\alpha_j-1}$ polarization layer and are sufficient for the time-integrated electromagnetic power balance.
Indeed, Eq.~\eqref{eq:cole_cole_constitutive} gives
\begin{equation}
{}^{C}_{0}D_t^{\alpha_j}\mathbf{P}_{j}
=
\frac{
\epsilon_0\boldsymbol{\chi}_{j}\mathbf{E}
-
\mathbf{P}_{j}}
{\tau_j^{\alpha_j}}
\in
L^{\infty}
\bigl(0,T;L^{2}(\Omega)^{3}\bigr).
\end{equation}
Together with
$\partial_t\mathbf{P}_{j}
=\mathcal{O}(t^{\alpha_j-1})$,
this implies
\begin{equation}
t\longmapsto
\int_{\Omega}
\partial_t\mathbf{P}_{j}
\cdot
\boldsymbol{\chi}_{j}^{-1}
{}^{C}_{0}D_t^{\alpha_j}\mathbf{P}_{j}\,dV
\in
L^{1}(0,T).
\end{equation}
The local energy identities in
Sec.~\ref{sec:energy_decomposition}
hold for almost every $t>0$. The global energy theorem is interpreted in time-integrated form. Calculations can first be performed on $[\delta,T]$, with $\delta>0$, and extended to $[0,T]$ by taking $\delta\to0^{+}$ and using Eq.~\eqref{eq:weighted_fractional_class}. This limit does not require a bounded instantaneous fractional dissipation rate at initialization.
For an open system, the electromagnetic boundary supply is retained through the Poynting trace pairing. For regular traces, the standard Maxwell boundary pairing reduces to the Poynting flux \cite{Monk2003},
\begin{equation}
\int_{\partial\Omega}
(\mathbf{E}\times\mathbf{H})
\cdot\mathbf{n}\,dA.
\end{equation}
A closed perfectly conducting system satisfies \cite{Jackson:1999,Monk2003}
\begin{equation}
\mathbf{n}\times\mathbf{E}
=
\mathbf{0}
\qquad
\text{on }\partial\Omega,
\label{eq:pec_boundary_condition}
\end{equation}
in the tangential trace sense of
$H(\operatorname{curl};\Omega)$, and the boundary power vanishes. Periodic conditions give the same cancellation after integration over $\Omega$. The uniqueness argument in
Sec.~\ref{sec:global_energy}
is restricted to PEC, periodic, or another homogeneous passive boundary condition for which the difference of two solutions has zero boundary supply.
Equations~\eqref{eq:maxwell_system}--\eqref{eq:pec_boundary_condition} define a linear, anisotropic, multichannel Maxwell--Cole--Cole initial-value problem with zero prehistory, passive conduction, and weakly singular fractional relaxation. The isotropic single-channel model follows from
$N=1$,
$\boldsymbol{\epsilon}_{\infty}=\epsilon_{\infty}\mathbf{I}$,
$\boldsymbol{\chi}_{1}=\chi_{1}\mathbf{I}$, and
$\boldsymbol{\sigma}=\sigma\mathbf{I}$.
The energy decomposition derived below follows directly from the Caputo convolution and does not require a generalized fractional Leibniz rule.

\section{Energy balance and history decomposition}
\label{sec:energy_decomposition}
The energy identities are derived for smooth fields and then extended to the weak solution class of Sec.~\ref{sec:fractional_model}. Time derivatives are understood for almost every $t\in(0,T)$. Singular history integrals are evaluated on truncated intervals, and their endpoints are recovered by dominated convergence.

Define the Poynting vector by $\mathbf{S}=\mathbf{E}\times\mathbf{H}$. Using the standard vector identity and the macroscopic Maxwell equations, the Poynting balance is obtained as follows \cite{Jackson:1999}. The identity
\begin{equation}
\nabla\cdot(\mathbf{E}\times\mathbf{H})
=
\mathbf{H}\cdot(\nabla\times\mathbf{E})
-
\mathbf{E}\cdot(\nabla\times\mathbf{H})
\end{equation}
and the Maxwell curl equations give
\begin{equation}
-\nabla\cdot\mathbf{S}
=
\mathbf{E}\cdot\mathbf{J}_{f}
+
\mathbf{E}\cdot\partial_t\mathbf{D}
+
\mathbf{H}\cdot\partial_t\mathbf{B}.
\end{equation}
This identity is independent of the dielectric constitutive model.
Since $\boldsymbol{\epsilon}_{\infty}$ is symmetric and time independent,
\begin{equation}
\mathbf{E}\cdot\partial_t\mathbf{D}
+
\mathbf{H}\cdot\partial_t\mathbf{B}
=
\partial_t
\left[
\frac{\epsilon_0}{2}
\mathbf{E}\cdot
\boldsymbol{\epsilon}_{\infty}\mathbf{E}
+
\frac{\mu_0}{2}
|\mathbf{H}|^{2}
\right]
+
\sum_{j=1}^{N}
\mathbf{E}\cdot\partial_t\mathbf{P}_{j}.
\end{equation}
The constitutive law in Eq.~\eqref{eq:cole_cole_constitutive} may be written as
\begin{equation}
\mathbf{E}
=
\frac{1}{\epsilon_0}
\boldsymbol{\chi}_{j}^{-1}
\left(
\mathbf{P}_{j}
+
\tau_j^{\alpha_j}
{}^{C}_{0}D_t^{\alpha_j}\mathbf{P}_{j}
\right).
\end{equation}
Taking its scalar product with $\partial_t\mathbf{P}_{j}$ yields
\begin{equation}
\mathbf{E}\cdot\partial_t\mathbf{P}_{j}
=
\partial_t
\left[
\frac{1}{2\epsilon_0}
\left|
\mathbf{P}_{j}
\right|_{\boldsymbol{\chi}_{j}^{-1}}^{2}
\right]
+
F_{\alpha_j},
\end{equation}
where
\begin{equation}
F_{\alpha_j}
=
\frac{\tau_j^{\alpha_j}}{\epsilon_0}
\partial_t\mathbf{P}_{j}
\cdot
\boldsymbol{\chi}_{j}^{-1}
{}^{C}_{0}D_t^{\alpha_j}\mathbf{P}_{j}
\label{eq:fractional_power_density}
\end{equation}
is the fractional polarization power density whenever the pointwise product is defined. Under the weak spatial regularity of Sec.~\ref{sec:fractional_model}, its spatially integrated form is the primary object.
Define the instantaneous field and polarization energy density by
\begin{equation}
u_{\mathrm{inst}}
=
\frac{\epsilon_0}{2}
\mathbf{E}\cdot
\boldsymbol{\epsilon}_{\infty}\mathbf{E}
+
\frac{\mu_0}{2}
|\mathbf{H}|^{2}
+
\sum_{j=1}^{N}
\frac{1}{2\epsilon_0}
\left|
\mathbf{P}_{j}
\right|_{\boldsymbol{\chi}_{j}^{-1}}^{2}.
\label{eq:instantaneous_energy_density}
\end{equation}
Using
$\mathbf{J}_{f}=\boldsymbol{\sigma}\mathbf{E}+\mathbf{J}_{\mathrm{ext}}$
then gives
\begin{equation}
-\nabla\cdot\mathbf{S}
=
\partial_tu_{\mathrm{inst}}
+
\mathbf{E}\cdot\boldsymbol{\sigma}\mathbf{E}
+
\mathbf{E}\cdot\mathbf{J}_{\mathrm{ext}}
+
\sum_{j=1}^{N}F_{\alpha_j}.
\label{eq:poynting_balance_with_fractional_power}
\end{equation}
Each $F_{\alpha_j}$ is next decomposed into history-dependent storage and non-negative constitutive dissipation.
For channel $j$, introduce the weighted Hilbert-space inner product
\begin{subequations}
\label{eq:weighted_history_inner_product_and_norm}
\begin{align}
\langle\mathbf{u},\mathbf{v}\rangle_{j,\Omega}
&=
\int_{\Omega}
\mathbf{u}\cdot
\boldsymbol{\chi}_{j}^{-1}\mathbf{v}\,dV,
\label{eq:weighted_history_inner_product}
\\
|\mathbf{u}|_{j,\Omega}^{2}
&=
\langle\mathbf{u},\mathbf{u}\rangle_{j,\Omega}.
\label{eq:weighted_history_norm}
\end{align}
\end{subequations}
Uniform positive definiteness of $\boldsymbol{\chi}_{j}$ makes this norm equivalent to the usual $L^{2}(\Omega)^{3}$ norm. The integrated fractional power is
\begin{equation}
\mathcal{F}_{\alpha_j}(t)
=
\frac{\tau_j^{\alpha_j}}{\epsilon_0}
\left\langle
\partial_t\mathbf{P}_{j}(t),
{}^{C}_{0}D_t^{\alpha_j}\mathbf{P}_{j}(t)
\right\rangle_{j,\Omega}.
\label{eq:integrated_fractional_power}
\end{equation}
The regularity assumptions in Sec.~\ref{sec:fractional_model} imply
$\mathcal{F}_{\alpha_j}\in L^{1}(0,T)$.
For $0\leq s\leq t\leq T$, set
\begin{equation}
\mathcal{G}_{j}(t,s)
=
\left|
\mathbf{P}_{j}(t)-\mathbf{P}_{j}(s)
\right|_{j,\Omega}^{2}.
\end{equation}
Define the history-storage functional and its associated dissipation rate by
\begin{align}
\mathcal{U}_{\alpha_j}(t)
&=
\frac{\tau_j^{\alpha_j}}
{2\epsilon_0\Gamma(1-\alpha_j)}
\Bigg[
t^{-\alpha_j}
\left|
\mathbf{P}_{j}(t)
\right|_{j,\Omega}^{2}
\nonumber\\
&\hspace{2.3cm}
+
\alpha_j
\int_{0}^{t}
\frac{
\mathcal{G}_{j}(t,s)}
{(t-s)^{\alpha_j+1}}
\,ds
\Bigg],
\label{eq:integrated_history_energy}
\\
\mathcal{Q}_{\alpha_j}(t)
&=
\frac{
\tau_j^{\alpha_j}\alpha_j}
{2\epsilon_0\Gamma(1-\alpha_j)}
\Bigg[
t^{-\alpha_j-1}
\left|
\mathbf{P}_{j}(t)
\right|_{j,\Omega}^{2}
\nonumber\\
&\hspace{1.9cm}
+
(\alpha_j+1)
\int_{0}^{t}
\frac{
\mathcal{G}_{j}(t,s)}
{(t-s)^{\alpha_j+2}}
\,ds
\Bigg].
\label{eq:integrated_history_dissipation}
\end{align}
The resulting identity is
\begin{equation}
\mathcal{F}_{\alpha_j}(t)
=
\frac{d\mathcal{U}_{\alpha_j}}{dt}
+
\mathcal{Q}_{\alpha_j}(t)
\label{eq:integrated_history_split}
\end{equation}
for almost every $t\in(0,T)$, with
\begin{subequations}
\label{eq:integrated_history_positivity}
\begin{align}
\mathcal{U}_{\alpha_j}(t)&\geq0,
\label{eq:integrated_history_storage_positivity}
\\
\mathcal{Q}_{\alpha_j}(t)&\geq0.
\label{eq:integrated_history_dissipation_positivity}
\end{align}
\end{subequations}
In addition,
$\mathcal{U}_{\alpha_j}\in AC([0,T])$,
$\mathcal{U}_{\alpha_j}(0)=0$, and
$\mathcal{Q}_{\alpha_j}\in L^{1}(0,T)$.
To prove Eq.~\eqref{eq:integrated_history_split}, substitute the Caputo convolution into Eq.~\eqref{eq:integrated_fractional_power} to obtain
\begin{equation}
\mathcal{F}_{\alpha_j}(t)
=
\frac{\tau_j^{\alpha_j}}
{\epsilon_0\Gamma(1-\alpha_j)}
\int_{0}^{t}
\frac{
\langle
\partial_t\mathbf{P}_{j}(t),
\partial_t\mathbf{P}_{j}(s)
\rangle_{j,\Omega}}
{(t-s)^{\alpha_j}}
\,ds.
\end{equation}
For $s,t>0$, Hilbert-space differentiation gives
\begin{equation}
\partial_t\partial_s\mathcal{G}_j(t,s)
=
-2
\left\langle
\partial_t\mathbf{P}_j(t),
\partial_t\mathbf{P}_j(s)
\right\rangle_{j,\Omega}.
\end{equation}
Hence,
\begin{equation}
\mathcal{F}_{\alpha_j}(t)
=
-
\frac{\tau_j^{\alpha_j}}
{2\epsilon_0\Gamma(1-\alpha_j)}
\int_0^t
\frac{
\partial_t\partial_s\mathcal{G}_j(t,s)}
{(t-s)^{\alpha_j}}
\,ds.
\end{equation}

For $0<\varepsilon<t-\eta<t$, integration by parts over
$[\varepsilon,t-\eta]$ gives
\begin{align}
\mathcal{F}_{\alpha_j}^{\varepsilon,\eta}(t)
&=
-
\frac{\tau_j^{\alpha_j}}
{2\epsilon_0\Gamma(1-\alpha_j)}
\left[
\frac{
\partial_t\mathcal{G}_j(t,s)}
{(t-s)^{\alpha_j}}
\right]_{s=\varepsilon}^{s=t-\eta}
\nonumber\\
&\quad+
\frac{\tau_j^{\alpha_j}\alpha_j}
{2\epsilon_0\Gamma(1-\alpha_j)}
\int_{\varepsilon}^{t-\eta}
\frac{
\partial_t\mathcal{G}_j(t,s)}
{(t-s)^{\alpha_j+1}}
\,ds.
\end{align}
Since $\mathbf{P}_{j}$ is continuously differentiable on compact subintervals of $(0,T]$,
\begin{equation}
\mathbf{P}_{j}(t)-\mathbf{P}_{j}(s)
=
(t-s)\partial_t\mathbf{P}_{j}(t)
+
o_{L^{2}}(t-s)
\end{equation}
as $s\to t^{-}$. Therefore,
\begin{subequations}
\label{eq:diagonal_increment_estimates}
\begin{align}
\mathcal{G}_{j}(t,s)
&=
\mathcal{O}\bigl((t-s)^{2}\bigr),
\label{eq:diagonal_increment_estimate}
\\
\partial_t\mathcal{G}_{j}(t,s)
&=
\mathcal{O}(t-s).
\label{eq:diagonal_increment_derivative_estimate}
\end{align}
\end{subequations}
The upper boundary term tends to zero because
$(t-s)^{-\alpha_j}\partial_t\mathcal{G}_{j}
=
\mathcal{O}((t-s)^{1-\alpha_j})$.

At the lower endpoint,
$\partial_t\mathbf{P}_{j}\in
L^{1}(0,T;L^{2}(\Omega)^{3})$
and
$\mathbf{P}_{j}(s)\to\mathbf{P}_{j}(0)=\mathbf{0}$.
Consequently,
\begin{equation}
\partial_t\mathcal{G}_{j}(t,0)
=
\frac{d}{dt}
\left|
\mathbf{P}_{j}(t)
\right|_{j,\Omega}^{2}.
\end{equation}
Taking $\varepsilon,\eta\to0^{+}$ yields
\begin{equation}
\begin{aligned}
\mathcal{F}_{\alpha_j}(t)
={}&
\frac{\tau_j^{\alpha_j}}
{2\epsilon_0\Gamma(1-\alpha_j)}
\Bigg[
t^{-\alpha_j}
\frac{d}{dt}
\left|
\mathbf{P}_{j}(t)
\right|_{j,\Omega}^{2}
\\
&+
\alpha_j
\int_{0}^{t}
\frac{
\partial_t\mathcal{G}_{j}(t,s)}
{(t-s)^{\alpha_j+1}}
\,ds
\Bigg].
\end{aligned}
\end{equation}
To differentiate $\mathcal{U}_{\alpha_j}$, its history integral is first truncated at $t-\eta$. The estimates
\begin{subequations}
\label{eq:differentiated_kernel_estimates}
\begin{align}
\frac{
\partial_t\mathcal{G}_{j}(t,s)}
{(t-s)^{\alpha_j+1}}
&=
\mathcal{O}\bigl((t-s)^{-\alpha_j}\bigr),
\label{eq:differentiated_increment_kernel_estimate}
\\
\frac{
\mathcal{G}_{j}(t,s)}
{(t-s)^{\alpha_j+2}}
&=
\mathcal{O}\bigl((t-s)^{-\alpha_j}\bigr).
\label{eq:differentiated_history_kernel_estimate}
\end{align}
\end{subequations}
show that the differentiated kernels are integrable near $s=t$ because
$0<\alpha_j<1$. The ordinary Leibniz rule on the truncated interval, followed by dominated convergence, gives
\begin{equation}
\begin{aligned}
\frac{d\mathcal{U}_{\alpha_j}}{dt}
={}&
\frac{\tau_j^{\alpha_j}}
{2\epsilon_0\Gamma(1-\alpha_j)}
\Bigg[
t^{-\alpha_j}
\frac{d}{dt}
\left|
\mathbf{P}_{j}(t)
\right|_{j,\Omega}^{2}
\\
&+
\alpha_j
\int_{0}^{t}
\frac{
\partial_t\mathcal{G}_{j}(t,s)}
{(t-s)^{\alpha_j+1}}
\,ds
\\
&-
\alpha_j t^{-\alpha_j-1}
\left|
\mathbf{P}_{j}(t)
\right|_{j,\Omega}^{2}
\\
&-
\alpha_j(\alpha_j+1)
\int_{0}^{t}
\frac{
\mathcal{G}_{j}(t,s)}
{(t-s)^{\alpha_j+2}}
\,ds
\Bigg].
\end{aligned}
\end{equation}
Comparison with the preceding expression for
$\mathcal{F}_{\alpha_j}$ proves
Eq.~\eqref{eq:integrated_history_split}. The positivity in
Eq.~\eqref{eq:integrated_history_positivity} follows directly from
$\mathcal{G}_{j}(t,s)\geq0$ and the positive material coefficients.
The behavior at initialization follows from the weighted temporal class. Let
\begin{equation}
M_{\alpha_j}
=
\sup_{0<t\leq T}
t^{1-\alpha_j}
\left|
\partial_t\mathbf{P}_{j}(t)
\right|_{j,\Omega}
<\infty.
\end{equation}
The quiescent initial condition gives
\begin{subequations}
\label{eq:polarization_initial_bounds}
\begin{align}
\left|
\mathbf{P}_{j}(t)
\right|_{j,\Omega}
&\leq
\frac{M_{\alpha_j}}{\alpha_j}t^{\alpha_j},
\label{eq:polarization_initial_bound}
\\
\left|
\mathbf{P}_{j}(t)-\mathbf{P}_{j}(s)
\right|_{j,\Omega}
&\leq
\frac{M_{\alpha_j}}{\alpha_j}
\left(t^{\alpha_j}-s^{\alpha_j}\right).
\label{eq:polarization_increment_initial_bound}
\end{align}
\end{subequations}
After setting $s=tx$, the two history terms are controlled by
\begin{subequations}
\label{eq:history_integrability_bounds}
\begin{align}
\int_{0}^{1}
\frac{(1-x^{\alpha_j})^{2}}
{(1-x)^{\alpha_j+1}}\,dx
&<\infty,
\label{eq:history_storage_integrability_bound}
\\
\int_{0}^{1}
\frac{(1-x^{\alpha_j})^{2}}
{(1-x)^{\alpha_j+2}}\,dx
&<\infty.
\label{eq:history_dissipation_integrability_bound}
\end{align}
\end{subequations}
Their integrands behave as
$(1-x)^{1-\alpha_j}$ and
$(1-x)^{-\alpha_j}$ near $x=1$, respectively. Hence
\begin{subequations}
\label{eq:history_initial_scaling_bounds}
\begin{align}
0
\leq
\mathcal{U}_{\alpha_j}(t)
&\leq
C_{\alpha_j}^{(U)}t^{\alpha_j},
\label{eq:history_storage_initial_scaling_bound}
\\
0
\leq
\mathcal{Q}_{\alpha_j}(t)
&\leq
C_{\alpha_j}^{(Q)}t^{\alpha_j-1}.
\label{eq:history_dissipation_initial_scaling_bound}
\end{align}
\end{subequations}
where the constants are independent of $t$. It follows that
$\mathcal{U}_{\alpha_j}(t)\to0$ as $t\to0^{+}$ and
$\mathcal{Q}_{\alpha_j}\in L^{1}(0,T)$. The instantaneous dissipation may diverge as
$t^{\alpha_j-1}$ for nonzero initial electric data, but the divergence is integrable.
Since
$\mathcal{F}_{\alpha_j},\mathcal{Q}_{\alpha_j}\in L^{1}(0,T)$,
Eq.~\eqref{eq:integrated_history_split} implies
$\mathcal{U}_{\alpha_j}\in AC([0,T])$. Integration from $0$ to $t$ gives the channelwise passivity inequality
\begin{equation}
\int_{0}^{t}
\mathcal{F}_{\alpha_j}(\eta)\,d\eta
=
\mathcal{U}_{\alpha_j}(t)
+
\int_{0}^{t}
\mathcal{Q}_{\alpha_j}(\eta)\,d\eta
\geq0.
\label{eq:fractional_material_passivity}
\end{equation}
Thus, the cumulative fractional work supplied through a polarization channel cannot be negative when its initial memory vanishes. A nonzero prehistory would add the change of the corresponding initial storage. For fixed $t>0$, either
$\mathcal{U}_{\alpha_j}(t)=0$ or
$\mathcal{Q}_{\alpha_j}(t)=0$
implies
$\mathbf{P}_{j}(s)=\mathbf{0}$ for every $s\in[0,t]$.
The Hilbert-space statement applies under the weak spatial assumptions of Sec.~\ref{sec:fractional_model}. If, in addition,
$\mathbf{P}_{j}(\mathbf{r},\cdot)
\in\mathcal{X}_{\alpha_j}(0,T;\mathbb{R}^{3})$
for almost every $\mathbf{r}\in\Omega$, with spatially integrable weighted bounds, define
\begin{equation}
G_j(\mathbf{r};t,s)
=
\left|
\mathbf{P}_{j}(\mathbf{r},t)
-
\mathbf{P}_{j}(\mathbf{r},s)
\right|_{\boldsymbol{\chi}_{j}^{-1}}^{2}.
\end{equation}
The local history-storage and dissipation densities are
\begin{align}
U_{\alpha_j}(\mathbf{r},t)
&=
\frac{\tau_j^{\alpha_j}}
{2\epsilon_0\Gamma(1-\alpha_j)}
\Bigg[
t^{-\alpha_j}
\left|
\mathbf{P}_{j}(\mathbf{r},t)
\right|_{\boldsymbol{\chi}_{j}^{-1}}^{2}
\nonumber\\
&\hspace{2.1cm}
+
\alpha_j
\int_{0}^{t}
\frac{
G_j(\mathbf{r};t,s)}
{(t-s)^{\alpha_j+1}}
\,ds
\Bigg],
\label{eq:history_energy_functional}
\\
Q_{\alpha_j}(\mathbf{r},t)
&=
\frac{
\tau_j^{\alpha_j}\alpha_j}
{2\epsilon_0\Gamma(1-\alpha_j)}
\Bigg[
t^{-\alpha_j-1}
\left|
\mathbf{P}_{j}(\mathbf{r},t)
\right|_{\boldsymbol{\chi}_{j}^{-1}}^{2}
\nonumber\\
&\hspace{1.7cm}
+
(\alpha_j+1)
\int_{0}^{t}
\frac{
G_j(\mathbf{r};t,s)}
{(t-s)^{\alpha_j+2}}
\,ds
\Bigg].
\label{eq:history_dissipation_functional}
\end{align}
They satisfy
\begin{subequations}
\label{eq:history_split}
\begin{align}
F_{\alpha_j}
&=
\partial_tU_{\alpha_j}
+
Q_{\alpha_j},
\label{eq:local_history_power_split}
\\
U_{\alpha_j}&\geq0,
\label{eq:local_history_storage_positivity}
\\
Q_{\alpha_j}&\geq0.
\label{eq:local_history_dissipation_positivity}
\end{align}
\end{subequations}
for almost every
$(\mathbf{r},t)\in\Omega\times(0,T)$, and
\begin{subequations}
\label{eq:spatial_history_functionals}
\begin{align}
\mathcal{U}_{\alpha_j}(t)
&=
\int_{\Omega}U_{\alpha_j}(\mathbf{r},t)\,dV,
\label{eq:spatial_history_storage}
\\
\mathcal{Q}_{\alpha_j}(t)
&=
\int_{\Omega}Q_{\alpha_j}(\mathbf{r},t)\,dV.
\label{eq:spatial_history_dissipation}
\end{align}
\end{subequations}
Define the total local storage and material dissipation by
\begin{align}
u_{\mathrm{tot}}
&=
u_{\mathrm{inst}}
+
\sum_{j=1}^{N}U_{\alpha_j},
\label{eq:total_local_energy_density}
\\
q_{\mathrm{diss}}
&=
\mathbf{E}\cdot\boldsymbol{\sigma}\mathbf{E}
+
\sum_{j=1}^{N}Q_{\alpha_j}.
\label{eq:total_material_dissipation}
\end{align}
Substitution into
Eq.~\eqref{eq:poynting_balance_with_fractional_power}
gives the completed local energy balance
\begin{subequations}
\label{eq:completed_local_energy_balance}
\begin{align}
\partial_tu_{\mathrm{tot}}
+
\nabla\cdot\mathbf{S}
+
q_{\mathrm{diss}}
&=
-\mathbf{E}\cdot\mathbf{J}_{\mathrm{ext}},
\label{eq:local_total_energy_balance}
\\
q_{\mathrm{diss}}&\geq0.
\label{eq:local_total_dissipation_positivity}
\end{align}
\end{subequations}
in the distributional sense. The right-hand side is the work supplied to the electromagnetic system by the prescribed current, while inward Poynting flux supplies energy through the boundary.
Equation~\eqref{eq:completed_local_energy_balance} separates electromagnetic transport, instantaneous field and polarization energy, hereditary storage, conductive loss, fractional constitutive dissipation, and work by the external current. Related energy structures appear in stability analyses of time-domain Maxwell--Cole--Cole systems \cite{LiHuangLin2011,YinLiuLiZhang2023}.
The functional $U_{\alpha_j}$ is an admissible non-negative storage measure for the finite-terminal constitutive law. It establishes material passivity but does not define a unique thermodynamic free energy. Materials with memory can admit several storage or free-energy functionals for the same constitutive response \cite{Fabrizio:2014,DeseriDiPaolaZingales2014,AmendolaFabrizioGolden2021}. A thermodynamic interpretation would also require temperature evolution, heat transport, and an entropy inequality. Section~\ref{sec:diffusive_representation} constructs a second positive storage representation of the same fractional polarization power.

\section{Global energy theorem and limiting cases}
\label{sec:global_energy}
The local distributional balance in
Eq.~\eqref{eq:completed_local_energy_balance}
determines the energy evolution over $\Omega$. Under the weak regularity assumptions of Sec.~\ref{sec:fractional_model}, define the instantaneous energy, total field--memory storage, and passive dissipation rate by

\begin{subequations}
\label{eq:global_energy_functionals}
\begin{align}
\mathcal{E}_{\mathrm{inst}}(t)
&=
\int_{\Omega}
\left[
\frac{\epsilon_0}{2}
\mathbf{E}\cdot
\boldsymbol{\epsilon}_{\infty}\mathbf{E}
+
\frac{\mu_0}{2}
|\mathbf{H}|^{2}
+
\sum_{j=1}^{N}
\frac{1}{2\epsilon_0}
\left|
\mathbf{P}_{j}
\right|_{\boldsymbol{\chi}_{j}^{-1}}^{2}
\right]\,dV,
\label{eq:global_instantaneous_energy}
\\
\mathcal{E}(t)
&=
\mathcal{E}_{\mathrm{inst}}(t)
+
\sum_{j=1}^{N}
\mathcal{U}_{\alpha_j}(t),
\label{eq:global_total_energy}
\\
\mathcal{D}(t)
&=
\int_{\Omega}
\mathbf{E}\cdot
\boldsymbol{\sigma}\mathbf{E}\,dV
+
\sum_{j=1}^{N}
\mathcal{Q}_{\alpha_j}(t).
\label{eq:global_total_dissipation}
\end{align}
\end{subequations}
Here, $\mathcal{U}_{\alpha_j}$ and
$\mathcal{Q}_{\alpha_j}$ are the history-storage and dissipation functionals defined in
Eqs.~\eqref{eq:integrated_history_energy} and
\eqref{eq:integrated_history_dissipation}. The positivity of the material tensors and
Eq.~\eqref{eq:integrated_history_positivity} imply
\begin{subequations}
\label{eq:global_energy_positivity}
\begin{align}
\mathcal{E}(t)&\geq0,
\label{eq:global_storage_positivity}
\\
\mathcal{D}(t)&\geq0
\quad
\text{for almost every }t\in(0,T).
\label{eq:global_dissipation_positivity}
\end{align}
\end{subequations}
Moreover,
$\mathcal{E}\in AC([0,T])$ and
$\mathcal{D}\in L^{1}(0,T)$.
Let $\Phi_S(t)$ denote the outward electromagnetic boundary power, and let
$\mathcal{W}_{\mathrm{ext}}(t)$ denote the work rate transferred from the field to the prescribed current. Define
\begin{subequations}
\label{eq:global_supply_terms}
\begin{align}
\Phi_S(t)
&=
\int_{\partial\Omega}
(\mathbf{E}\times\mathbf{H})
\cdot\mathbf{n}\,dA,
\label{eq:boundary_poynting_flux}
\\
\mathcal{W}_{\mathrm{ext}}(t)
&=
\int_{\Omega}
\mathbf{E}\cdot
\mathbf{J}_{\mathrm{ext}}\,dV,
\label{eq:external_current_work}
\\
\mathcal{P}_{\mathrm{in}}(t)
&=
-\Phi_S(t)
-
\mathcal{W}_{\mathrm{ext}}(t).
\label{eq:net_input_power}
\end{align}
\end{subequations}
For weak $H(\operatorname{curl};\Omega)$ fields,
Eq.~\eqref{eq:boundary_poynting_flux}
denotes the corresponding tangential-trace duality pairing \cite{Monk2003}. Positive $\Phi_S$ represents energy leaving $\Omega$, whereas positive
$\mathcal{W}_{\mathrm{ext}}$ represents work done by the electromagnetic field on the imposed current.
Integration of
Eq.~\eqref{eq:completed_local_energy_balance}
over $\Omega\times[t_1,t_2]$, where
$0\leq t_1\leq t_2\leq T$, gives
\begin{subequations}
\label{eq:global_energy_theorem}
\begin{align}
\mathcal{E}(t_2)
+
\int_{t_1}^{t_2}
\mathcal{D}(t)\,dt
&=
\mathcal{E}(t_1)
-
\int_{t_1}^{t_2}
\Phi_S(t)\,dt
-
\int_{t_1}^{t_2}
\mathcal{W}_{\mathrm{ext}}(t)\,dt,
\label{eq:integrated_global_energy_balance}
\\
\mathcal{E}(t_2)
-
\mathcal{E}(t_1)
+
\int_{t_1}^{t_2}
\mathcal{D}(t)\,dt
&=
\int_{t_1}^{t_2}
\mathcal{P}_{\mathrm{in}}(t)\,dt.
\label{eq:integrated_global_supply_balance}
\end{align}
\end{subequations}
Since $\mathcal{E}$ is absolutely continuous,
\begin{equation}
\frac{d\mathcal{E}}{dt}
+
\mathcal{D}(t)
=
\mathcal{P}_{\mathrm{in}}(t)
\label{eq:global_energy_balance}
\end{equation}
for almost every $t\in(0,T)$. The integrated identity
\eqref{eq:integrated_global_supply_balance}
is the primary theorem for solutions containing the fractional initial layer described in Sec.~\ref{sec:fractional_model}. Related energy estimates for Maxwell--Cole--Cole systems are given in
Refs.~\cite{LiHuangLin2011,YinLiuLiZhang2023}.
Dropping the non-negative accumulated dissipation gives
\begin{equation}
\mathcal{E}(t_2)-\mathcal{E}(t_1)
\leq
\int_{t_1}^{t_2}
\mathcal{P}_{\mathrm{in}}(t)\,dt.
\label{eq:global_passivity_inequality}
\end{equation}
Thus, $\mathcal{E}$ is a non-negative storage functional for the boundary and external-current supply. In particular, zero initial fields and zero initial memory imply
\begin{equation}
\mathcal{E}(t)
\leq
\int_{0}^{t}
\mathcal{P}_{\mathrm{in}}(\eta)\,d\eta.
\end{equation}
The channelwise constitutive counterpart is
Eq.~\eqref{eq:fractional_material_passivity}.
For the PEC condition in
Eq.~\eqref{eq:pec_boundary_condition},
the tangential electric trace vanishes and hence
$\Phi_S=0$. Periodic conditions give the same cancellation after integration over $\Omega$.
More generally, the equalities below require a homogeneous boundary condition with zero boundary power, $\Phi_S=0$, and $\mathbf{J}_{\mathrm{ext}}=\mathbf{0}$. A dissipative passive boundary instead has $\Phi_S\geq0$ and adds a non-negative boundary-loss term to the balance. Under the zero-boundary-power assumption,
\begin{subequations}
\label{eq:closed_system_decay}
\begin{align}
\mathcal{E}(t_2)
+
\int_{t_1}^{t_2}
\mathcal{D}(t)\,dt
&=
\mathcal{E}(t_1),
\label{eq:closed_system_integrated_decay}
\\
\frac{d\mathcal{E}}{dt}
&=
-\mathcal{D}(t)
\leq0.
\label{eq:closed_system_energy_decay}
\end{align}
\end{subequations}
The total field--memory energy is therefore non-increasing, and its decrease equals the conductive and fractional constitutive losses.
The same estimate gives conditional uniqueness. Let two weak solutions have identical material tensors, prescribed current, divergence constraints, initial fields, polarization prehistory, and boundary data. If the difference problem has zero boundary supply, then
\begin{equation}
\mathcal{E}_{\Delta}(t)
+
\int_{0}^{t}
\mathcal{D}_{\Delta}(\eta)\,d\eta
= 0.
\end{equation}
Both terms are non-negative, so
$\mathcal{E}_{\Delta}(t)=0$. Uniform positive definiteness of
$\boldsymbol{\epsilon}_{\infty}$ and
$\boldsymbol{\chi}_{j}$ implies
\begin{subequations}
\label{eq:uniqueness_field_equalities}
\begin{align}
\mathbf{E}^{(1)}&=\mathbf{E}^{(2)},
\label{eq:uniqueness_electric_field}
\\
\mathbf{H}^{(1)}&=\mathbf{H}^{(2)},
\label{eq:uniqueness_magnetic_field}
\\
\mathbf{P}_{j}^{(1)}&=\mathbf{P}_{j}^{(2)}.
\label{eq:uniqueness_polarization}
\end{align}
\end{subequations}
almost everywhere in
$\Omega\times(0,T)$. This proves uniqueness within the stated solution class, conditional on existence.
The Debye limit must distinguish a constitutive functional evaluated on a fixed history from a convergent family of coupled Maxwell solutions. First, fix
\begin{subequations}
\label{eq:fixed_debye_history_assumptions}
\begin{align}
\mathbf{P}_{j}
&\in
C^{1}\bigl([0,T];L^{2}(\Omega)^{3}\bigr),
\label{eq:fixed_debye_history_regularity}
\\
\mathbf{P}_{j}(0)&=\mathbf{0}.
\label{eq:fixed_debye_history_initial_value}
\end{align}
\end{subequations}
and a time $t>0$. Let
$\delta_j=1-\alpha_j\to0^{+}$, and define the backward time increment $h=t-s$ by
\begin{equation}
g_j(h;t)
=
\left\|
\mathbf{P}_{j}(t)
-
\mathbf{P}_{j}(t-h)
\right\|_{j,\Omega}^{2},
\qquad
0\leq h\leq t.
\label{eq:debye_history_increment}
\end{equation}
Differentiability at $t$ gives
\begin{equation}
\lim_{h\to0^{+}}
\frac{g_j(h;t)}{h^{2}}
=
\left\|
\partial_t\mathbf{P}_{j}(t)
\right\|_{j,\Omega}^{2}.
\label{eq:debye_history_local_limit}
\end{equation}
Since
$1/\Gamma(\delta_j)
=
\delta_j+\mathcal{O}(\delta_j^{2})$,
the bracket in
Eq.~\eqref{eq:integrated_history_energy}
remains bounded for the fixed history, and therefore
\begin{equation}
\lim_{\alpha_j\to1^{-}}
\mathcal{U}_{\alpha_j}(t)
= 0.
\label{eq:debye_history_energy_limit}
\end{equation}
The singular part of the dissipation is
\begin{equation}
\delta_j
\int_{0}^{t}
\frac{g_j(h;t)}{h^{3-\delta_j}}\,dh
=
\delta_j
\int_{0}^{t}
\frac{g_j(h;t)}{h^{2}}
h^{\delta_j-1}\,dh.
\end{equation}
Splitting the integral over
$(0,\eta)$ and $(\eta,t)$, using
Eq.~\eqref{eq:debye_history_local_limit}, and then taking
$\delta_j\to0^{+}$ followed by
$\eta\to0^{+}$ gives the Abelian limit
\begin{equation}
\lim_{\delta_j\to0^{+}}
\delta_j
\int_{0}^{t}
\frac{g_j(h;t)}{h^{3-\delta_j}}\,dh
=
\left\|
\partial_t\mathbf{P}_{j}(t)
\right\|_{j,\Omega}^{2}.
\label{eq:debye_abelian_limit}
\end{equation}
Since
$\tau_j^{\alpha_j}\to\tau_j$ and
$\alpha_j(\alpha_j+1)/2\to1$,
\begin{equation}
\lim_{\alpha_j\to1^{-}}
\mathcal{Q}_{\alpha_j}(t)
=
\frac{\tau_j}{\epsilon_0}
\left\|
\partial_t\mathbf{P}_{j}(t)
\right\|_{j,\Omega}^{2}.
\label{eq:debye_dissipation_limit}
\end{equation}
The hereditary storage vanishes, while the fractional dissipation approaches the local Debye loss.
For a family
$\mathbf{P}_{j}^{(\alpha_j)}$ of coupled solutions, the same limit requires convergence of the histories. A sufficient condition on every
$[t_{\ast},T]$, where $t_{\ast}>0$, is
\begin{equation}
\mathbf{P}_{j}^{(\alpha_j)}
\longrightarrow
\mathbf{P}_{j}^{D}
\quad
\text{in }
C^{1}\bigl([t_{\ast},T];L^{2}(\Omega)^{3}\bigr),
\end{equation}
together with uniform bounds permitting dominated convergence. Since
\begin{equation}
{}^{C}_{0}D_t^{\alpha_j}\mathbf{P}_{j}
=
I^{1-\alpha_j}\partial_t\mathbf{P}_{j},
\end{equation}
the Riemann--Liouville integral acts as an approximate identity as
$\alpha_j\to1^{-}$ under the stated convergence assumptions \cite{Mainardi2022}. Equation~\eqref{eq:cole_cole_constitutive}
then approaches the standard anisotropic Debye law,
\begin{equation}
\tau_j
\partial_t\mathbf{P}_{j}^{D}
+
\mathbf{P}_{j}^{D}
=
\epsilon_0
\boldsymbol{\chi}_{j}
\mathbf{E}^{D},
\label{eq:debye_constitutive_limit}
\end{equation}
Unless uniform compatibility conditions are imposed, the convergence is local in time away from $t=0$.
The instantaneous dielectric limit is obtained by taking
$\tau_j\to0^{+}$ at fixed
$0<\alpha_j<1$. Let
$\mathbf{E}^{(\tau_j)}$ and
$\mathbf{P}_{j}^{(\tau_j)}$ solve the coupled problem. Assume that, for every
$t_{\ast}>0$,
\begin{equation}
\mathbf{E}^{(\tau_j)}
\longrightarrow
\mathbf{E}^{(0)}
\quad
\text{in }
C\bigl([t_{\ast},T];L^{2}(\Omega)^{3}\bigr),
\end{equation}
that
${}^{C}_{0}D_t^{\alpha_j}
\mathbf{P}_{j}^{(\tau_j)}$
is uniformly bounded on this interval, and that the unscaled history brackets are uniformly controlled. Equation~\eqref{eq:cole_cole_constitutive} then gives
\begin{equation}
\mathbf{P}_{j}^{(\tau_j)}
\longrightarrow
\epsilon_0
\boldsymbol{\chi}_{j}
\mathbf{E}^{(0)}
\quad
\text{in }
C\bigl([t_{\ast},T];L^{2}(\Omega)^{3}\bigr).
\label{eq:instantaneous_polarization_limit}
\end{equation}
The explicit factor $\tau_j^{\alpha_j}$ in the history functionals gives
\begin{subequations}
\label{eq:instantaneous_memory_limit}
\begin{align}
\mathcal{U}_{\alpha_j}^{(\tau_j)}(t)
&\longrightarrow0,
\label{eq:instantaneous_history_storage_limit}
\\
\mathcal{Q}_{\alpha_j}^{(\tau_j)}(t)
&\longrightarrow0,
\qquad t\geq t_{\ast}.
\label{eq:instantaneous_history_dissipation_limit}
\end{align}
\end{subequations}
The polarization-amplitude energy satisfies
\begin{equation}
\frac{1}{2\epsilon_0}
\left\|
\mathbf{P}_{j}^{(\tau_j)}(t)
\right\|_{\boldsymbol{\chi}_{j}^{-1}}^{2}
\longrightarrow
\frac{\epsilon_0}{2}
\int_{\Omega}
\mathbf{E}^{(0)}
\cdot
\boldsymbol{\chi}_{j}
\mathbf{E}^{(0)}\,dV.
\label{eq:instantaneous_polarization_energy}
\end{equation}
If all relaxation times tend to zero under the same assumptions, the total electric storage converges to
\begin{equation}
\mathcal{E}_{\mathrm{el}}^{(0)}(t)
=
\frac{\epsilon_0}{2}
\int_{\Omega}
\mathbf{E}^{(0)}
\cdot
\boldsymbol{\epsilon}_{s}
\mathbf{E}^{(0)}\,dV.
\label{eq:instantaneous_electric_energy}
\end{equation}
This convergence is generally not uniform at $t=0$. Quiescent polarization data with
$\mathbf{E}_{0}\neq\mathbf{0}$ produce a shrinking relaxation layer as
$\tau_j\to0^{+}$. Although
$\mathcal{Q}_{\alpha_j}^{(\tau_j)}(t)\to0$
for every fixed $t>0$, its accumulated contribution within the initial layer need not vanish. A global-in-time limit without an initial defect requires well-prepared data,
\begin{equation}
\mathbf{P}_{j}(0)
=
\epsilon_0
\boldsymbol{\chi}_{j}\mathbf{E}_{0},
\end{equation}
or, within the quiescent-data model,
$\mathbf{E}_{0}=\mathbf{0}$ together with compatible forcing near initialization.
For the vanishing-susceptibility limit, let
\begin{subequations}
\label{eq:vanishing_susceptibility_scaling}
\begin{align}
\boldsymbol{\chi}_{j}^{(\lambda)}
&=
\lambda
\overline{\boldsymbol{\chi}}_{j},
\qquad
0<\lambda\leq1,
\label{eq:vanishing_susceptibility_family}
\\
\mathbf{P}_{j}^{(\lambda)}
&=
\lambda
\widetilde{\mathbf{P}}_{j}^{(\lambda)},
\label{eq:vanishing_polarization_factorization}
\end{align}
\end{subequations}
where
$\overline{\boldsymbol{\chi}}_{j}$
is fixed and uniformly positive definite. The rescaled polarization satisfies
\begin{equation}
\tau_j^{\alpha_j}
{}^{C}_{0}D_t^{\alpha_j}
\widetilde{\mathbf{P}}_{j}^{(\lambda)}
+
\widetilde{\mathbf{P}}_{j}^{(\lambda)}
=
\epsilon_0
\overline{\boldsymbol{\chi}}_{j}
\mathbf{E}^{(\lambda)}.
\end{equation}
If $\mathbf{E}^{(\lambda)}$ remains uniformly bounded in the solution class of Sec.~\ref{sec:fractional_model}, then
$\widetilde{\mathbf{P}}_{j}^{(\lambda)}$ and its history norms remain uniformly bounded. Since
$(\boldsymbol{\chi}_{j}^{(\lambda)})^{-1}
=
\lambda^{-1}
\overline{\boldsymbol{\chi}}_{j}^{-1}$,
\begin{subequations}
\begin{align}
\frac{1}{2\epsilon_0}
\left\|
\mathbf{P}_{j}^{(\lambda)}
\right\|_{
(\boldsymbol{\chi}_{j}^{(\lambda)})^{-1}}^{2}
&=
\mathcal{O}(\lambda),
\label{eq:vanishing_instantaneous_energy_scaling}
\\
\mathcal{U}_{\alpha_j}^{(\lambda)}(t)
&=
\mathcal{O}(\lambda),
\qquad
\mathcal{Q}_{\alpha_j}^{(\lambda)}(t)
=
\mathcal{O}(\lambda).
\label{eq:vanishing_memory_scaling}
\end{align}
\end{subequations}
For a prescribed electric field independent of $\lambda$, the polarization scaling is exact. In the coupled problem, only the order estimate follows unless the electric field is also unchanged.
Each channel therefore disappears continuously as its susceptibility tends to zero. Removing all channels and setting
$\boldsymbol{\epsilon}_{\infty}=\mathbf{I}$ and
$\boldsymbol{\sigma}=\mathbf{0}$ gives the standard vacuum electromagnetic energy \cite{Jackson:1999},
\begin{equation}
\mathcal{E}_{\mathrm{vac}}(t)
=
\int_{\Omega}
\left[
\frac{\epsilon_0}{2}
|\mathbf{E}|^{2}
+
\frac{\mu_0}{2}
|\mathbf{H}|^{2}
\right]\,dV,
\label{eq:vacuum_electromagnetic_energy}
\end{equation}
and Eq.~\eqref{eq:integrated_global_energy_balance}
reduces to the vacuum Poynting theorem.
The initialization behavior follows from
Eq.~\eqref{eq:polarization_initial_layer}. Suppose
\begin{equation}
\mathbf{E}(t)
=
\mathbf{E}_{0}
+
o_{L^{2}}(1)
\qquad
\text{as }t\to0^{+},
\end{equation}
and define
\begin{equation}
\mathbf{A}_{j}
=
\frac{
\epsilon_0
\boldsymbol{\chi}_{j}
\mathbf{E}_{0}}
{
\tau_j^{\alpha_j}
\Gamma(1+\alpha_j)}.
\label{eq:initial_polarization_coefficient}
\end{equation}
Then
\begin{subequations}
\label{eq:initial_polarization_expansion}
\begin{align}
\mathbf{P}_{j}(t)
&=
\mathbf{A}_{j}t^{\alpha_j}
+
o_{L^{2}}(t^{\alpha_j}),
\label{eq:initial_polarization_asymptotic_global}
\\
\partial_t\mathbf{P}_{j}(t)
&=
\alpha_j
\mathbf{A}_{j}t^{\alpha_j-1}
+
o_{L^{2}}(t^{\alpha_j-1}).
\label{eq:initial_polarization_derivative_global}
\end{align}
\end{subequations}
The weighted estimates of Sec.~\ref{sec:energy_decomposition} imply
$\mathcal{U}_{\alpha_j}(t)=\mathcal{O}(t^{\alpha_j})$
and
$\mathcal{Q}_{\alpha_j}(t)=\mathcal{O}(t^{\alpha_j-1})$.
Sharp leading coefficients require uniform control over the shrinking interval. Specifically,
\begin{subequations}
\label{eq:uniform_initial_expansion}
\begin{align}
\sup_{0<s\leq t}
\frac{
\left\|
\mathbf{P}_{j}(s)
-
\mathbf{A}_{j}s^{\alpha_j}
\right\|_{j,\Omega}}
{s^{\alpha_j}}
&\longrightarrow0,
\label{eq:uniform_initial_remainder}
\\
\sup_{0<s\leq t}
s^{1-\alpha_j}
\left\|
\partial_s\mathbf{P}_{j}(s)
-
\alpha_j
\mathbf{A}_{j}s^{\alpha_j-1}
\right\|_{j,\Omega}
&\longrightarrow0
\label{eq:uniform_initial_derivative_remainder}
\end{align}
\end{subequations}
as $t\to0^{+}$. These conditions follow from a uniform Volterra expansion under sufficient temporal regularity of the electric field.
Define
\begin{subequations}
\label{eq:initial_history_integrals}
\begin{align}
I_{\alpha}^{(U)}
&=
\int_{0}^{1}
\frac{
(1-x^{\alpha})^{2}}
{(1-x)^{\alpha+1}}
\,dx,
\label{eq:initial_history_energy_integral}
\\
I_{\alpha}^{(Q)}
&=
\int_{0}^{1}
\frac{
(1-x^{\alpha})^{2}}
{(1-x)^{\alpha+2}}
\,dx.
\label{eq:initial_history_dissipation_integral}
\end{align}
\end{subequations}
Both integrals are finite for
$0<\alpha<1$, because their integrands behave as
$(1-x)^{1-\alpha}$ and
$(1-x)^{-\alpha}$ near $x=1$.
Using $s=tx$, the uniform remainder estimates, and dominated convergence gives
\begin{subequations}
\label{eq:initial_history_asymptotic_pair}
\begin{align}
\mathcal{U}_{\alpha_j}(t)
&=
C_{\alpha_j,j}^{(U)}
\left\|
\mathbf{A}_{j}
\right\|_{j,\Omega}^{2}
t^{\alpha_j}
+
o(t^{\alpha_j}),
\label{eq:initial_history_energy_asymptotic}
\\
\mathcal{Q}_{\alpha_j}(t)
&=
C_{\alpha_j,j}^{(Q)}
\left\|
\mathbf{A}_{j}
\right\|_{j,\Omega}^{2}
t^{\alpha_j-1}
+
o(t^{\alpha_j-1}),
\label{eq:initial_history_dissipation_asymptotic}
\end{align}
\end{subequations}
where
\begin{subequations}
\label{eq:initial_history_coefficients}
\begin{align}
C_{\alpha_j,j}^{(U)}
&=
\frac{
\tau_j^{\alpha_j}}
{
2\epsilon_0
\Gamma(1-\alpha_j)}
\left[
1+
\alpha_j
I_{\alpha_j}^{(U)}
\right],
\label{eq:initial_history_energy_coefficient}
\\
C_{\alpha_j,j}^{(Q)}
&=
\frac{
\tau_j^{\alpha_j}
\alpha_j}
{
2\epsilon_0
\Gamma(1-\alpha_j)}
\left[
1+
(\alpha_j+1)
I_{\alpha_j}^{(Q)}
\right].
\label{eq:initial_history_dissipation_coefficient}
\end{align}
\end{subequations}
Both coefficients are positive. If
$\mathbf{E}_{0}\neq\mathbf{0}$,
uniform positive definiteness of
$\boldsymbol{\chi}_{j}$ gives
$\|\mathbf{A}_{j}\|_{j,\Omega}>0$. Hence
\begin{subequations}
\label{eq:correct_initial_memory_scaling}
\begin{align}
\mathcal{U}_{\alpha_j}(t)
&=
\mathcal{O}(t^{\alpha_j}),
\label{eq:initial_history_storage_scaling}
\\
\mathcal{Q}_{\alpha_j}(t)
&=
\mathcal{O}(t^{\alpha_j-1}).
\label{eq:initial_history_dissipation_scaling}
\end{align}
\end{subequations}
The history storage vanishes as $t^{\alpha_j}$, while the instantaneous fractional dissipation generally diverges as
$t^{\alpha_j-1}$. This divergence is integrable.
\begin{equation}
\int_{0}^{t}
\mathcal{Q}_{\alpha_j}(\eta)\,d\eta
=
\mathcal{O}(t^{\alpha_j})
\longrightarrow0
\qquad
\text{as }t\to0^{+}.
\label{eq:initial_accumulated_dissipation}
\end{equation}
Thus, no finite history storage or accumulated fractional loss is present at initialization, although the instantaneous dissipation rate need not be bounded.
Since
$\mathbf{P}_{j}(0)=\mathbf{0}$ and
$\mathcal{U}_{\alpha_j}(0)=0$, the initial total energy is
\begin{equation}
\mathcal{E}(0)
=
\int_{\Omega}
\left[
\frac{\epsilon_0}{2}
\mathbf{E}_{0}\cdot
\boldsymbol{\epsilon}_{\infty}\mathbf{E}_{0}
+
\frac{\mu_0}{2}
|\mathbf{H}_{0}|^{2}
\right]\,dV.
\label{eq:initial_total_energy}
\end{equation}
The global balance, closed-system decay, conditional uniqueness, and limiting cases are consistent with the history-storage construction. They establish electromagnetic and constitutive passivity for the finite-terminal, zero-prehistory, linear, spatially local model. The quantity $\mathcal{D}$ is an electromagnetic energy-loss rate, not an entropy-production rate. A thermodynamic interpretation also requires temperature evolution, heat transport, a positive absolute temperature, and a Clausius--Duhem inequality \cite{AmendolaFabrizioGolden2021}.

\section{Diffusive representation and comparison of memory energies}
\label{sec:diffusive_representation}
The history functional in Sec.~\ref{sec:energy_decomposition} describes material storage through increments between the present polarization and its past values. Following established positive diffusive representations of fractional operators \cite{Montseny1998,YuanAgrawal2002}, the same Caputo dynamics admit an infinite-state representation that resolves the power-law kernel into a positive continuum of exponential relaxation modes. Diffusive realizations convert the hereditary convolution into local first-order equations and support passive finite-dimensional approximations \cite{MonteghettiHaineMatignon2019,Diethelm2022}.

For channel $j$, let
\begin{equation}
K_{\alpha_j}(t)
=
\frac{t^{-\alpha_j}}{\Gamma(1-\alpha_j)},
\qquad
0<\alpha_j<1.
\end{equation}
The gamma reflection identity and the Laplace transform of
$\xi^{\alpha_j-1}$ give
\begin{subequations}
\label{eq:diffusive_spectral_representation}
\begin{align}
K_{\alpha_j}(t)
&=
\int_{0}^{\infty}
e^{-\xi t}
\mu_{\alpha_j}(\xi)\,d\xi,
\label{eq:kernel_spectral_representation}
\\
\mu_{\alpha_j}(\xi)
&=
\frac{\sin(\pi\alpha_j)}{\pi}
\xi^{\alpha_j-1},
\qquad
\xi>0.
\label{eq:diffusive_spectral_density}
\end{align}
\end{subequations}
The density $\mu_{\alpha_j}$ is positive, and the spectral variable $\xi$ has dimensions of inverse time.
For
$\mathbf{P}_{j}\in
\mathcal{X}_{\alpha_j}(0,T;L^{2}(\Omega)^{3})$,
define the diffusive state
\begin{equation}
\boldsymbol{\phi}_{j}(\xi,t)
=
\int_{0}^{t}
e^{-\xi(t-s)}
\partial_s\mathbf{P}_{j}(s)\,ds.
\label{eq:diffusive_variable}
\end{equation}
The integral is a Bochner integral because the weighted regularity in
Eq.~\eqref{eq:weighted_fractional_class} implies
$\partial_t\mathbf{P}_{j}\in
L^{1}(0,T;L^{2}(\Omega)^{3})$.
For each $\xi>0$,
\begin{subequations}
\label{eq:diffusive_state_system}
\begin{align}
\partial_t\boldsymbol{\phi}_{j}(\xi,t)
&=
-\xi\boldsymbol{\phi}_{j}(\xi,t)
+
\partial_t\mathbf{P}_{j}(t),
\label{eq:diffusive_variable_evolution}
\\
\boldsymbol{\phi}_{j}(\xi,0)
&=
\mathbf{0}.
\label{eq:diffusive_variable_initial_condition}
\end{align}
\end{subequations}
Thus, every spectral value defines a stable first-order mode driven by the polarization rate.
Substitution of
Eq.~\eqref{eq:kernel_spectral_representation}
into the Caputo convolution gives
\begin{equation}
{}^{C}_{0}D_t^{\alpha_j}\mathbf{P}_{j}(t)
=
\int_{0}^{\infty}
\mu_{\alpha_j}(\xi)
\boldsymbol{\phi}_{j}(\xi,t)\,d\xi.
\label{eq:caputo_diffusive_representation}
\end{equation}
The exchange of the time and spectral integrations follows from the Bochner--Fubini theorem because
\begin{equation}
\int_{0}^{t}
K_{\alpha_j}(t-s)
\left\|
\partial_s\mathbf{P}_{j}(s)
\right\|_{L^{2}(\Omega)}
\,ds
<
\infty
\end{equation}
for almost every $t>0$. The Cole--Cole constitutive law
\eqref{eq:cole_cole_constitutive} is therefore equivalent to
\begin{equation}
\tau_j^{\alpha_j}
\int_{0}^{\infty}
\mu_{\alpha_j}(\xi)
\boldsymbol{\phi}_{j}(\xi,t)\,d\xi
+
\mathbf{P}_{j}(t)
=
\epsilon_0
\boldsymbol{\chi}_{j}
\mathbf{E}(t),
\label{eq:diffusive_constitutive_law}
\end{equation}
together with
Eq.~\eqref{eq:diffusive_state_system}. This continuum realization is exact. Finite-dimensional models arise only after the spectral integral is approximated.
Introduce
\begin{subequations}
\label{eq:weighted_diffusive_inner_product_and_norm}
\begin{align}
\langle\mathbf{u},\mathbf{v}\rangle_{j,\Omega}
&=
\int_{\Omega}
\mathbf{u}\cdot
\boldsymbol{\chi}_{j}^{-1}
\mathbf{v}\,dV,
\label{eq:weighted_diffusive_inner_product}
\\
\|\mathbf{u}\|_{j,\Omega}^{2}
&=
\langle\mathbf{u},\mathbf{u}\rangle_{j,\Omega}.
\label{eq:weighted_diffusive_norm}
\end{align}
\end{subequations}
The diffusive storage and dissipation are
\begin{subequations}
\label{eq:integrated_diffusive_functionals}
\begin{align}
\mathcal{U}_{\alpha_j}^{\mathrm{diff}}(t)
&=
\frac{\tau_j^{\alpha_j}}
{2\epsilon_0}
\int_{0}^{\infty}
\mu_{\alpha_j}(\xi)
\left\|
\boldsymbol{\phi}_{j}(\xi,t)
\right\|_{j,\Omega}^{2}\,d\xi,
\label{eq:integrated_diffusive_energy}
\\
\mathcal{Q}_{\alpha_j}^{\mathrm{diff}}(t)
&=
\frac{\tau_j^{\alpha_j}}
{\epsilon_0}
\int_{0}^{\infty}
\xi\mu_{\alpha_j}(\xi)
\left\|
\boldsymbol{\phi}_{j}(\xi,t)
\right\|_{j,\Omega}^{2}\,d\xi.
\label{eq:integrated_diffusive_dissipation}
\end{align}
\end{subequations}
Both quantities are non-negative.
Their finiteness follows directly from the spectral asymptotics. As
$\xi\to0^{+}$,
\begin{equation}
\left\|
\boldsymbol{\phi}_{j}(\xi,t)
\right\|_{j,\Omega}
\leq
\int_{0}^{t}
\left\|
\partial_s\mathbf{P}_{j}(s)
\right\|_{j,\Omega}\,ds,
\end{equation}
so the energy and dissipation integrands behave as
$\xi^{\alpha_j-1}$ and
$\xi^{\alpha_j}$. For large $\xi$, splitting
Eq.~\eqref{eq:diffusive_variable}
at $t/2$ gives
\begin{equation}
\begin{aligned}
\left\|
\boldsymbol{\phi}_{j}(\xi,t)
\right\|_{j,\Omega}
\leq{}&
e^{-\xi t/2}
\int_{0}^{t/2}
\left\|
\partial_s\mathbf{P}_{j}(s)
\right\|_{j,\Omega}\,ds
\\
&+
\frac{1-e^{-\xi t/2}}{\xi}
\sup_{t/2\leq s\leq t}
\left\|
\partial_s\mathbf{P}_{j}(s)
\right\|_{j,\Omega}.
\end{aligned}
\end{equation}
Hence
$\|\boldsymbol{\phi}_{j}(\xi,t)\|_{j,\Omega}
=\mathcal{O}(\xi^{-1})$,
and the two integrands behave as
$\xi^{\alpha_j-3}$ and
$\xi^{\alpha_j-2}$ at infinity. These powers are integrable for
$0<\alpha_j<1$.
To derive the power identity without assuming regularity at $t=0$, truncate the spectral interval to
$[R^{-1},R]$. For $t\geq\delta>0$, differentiation under the finite integral and
Eq.~\eqref{eq:diffusive_variable_evolution} give
\begin{equation}
\begin{aligned}
\frac{d}{dt}
\mathcal{U}_{\alpha_j,R}^{\mathrm{diff}}
={}&
-\frac{\tau_j^{\alpha_j}}{\epsilon_0}
\int_{R^{-1}}^{R}
\xi\mu_{\alpha_j}(\xi)
\left\|
\boldsymbol{\phi}_{j}
\right\|_{j,\Omega}^{2}\,d\xi
\\
&+
\frac{\tau_j^{\alpha_j}}{\epsilon_0}
\left\langle
\partial_t\mathbf{P}_{j},
\int_{R^{-1}}^{R}
\mu_{\alpha_j}(\xi)
\boldsymbol{\phi}_{j}(\xi,t)\,d\xi
\right\rangle_{j,\Omega}.
\end{aligned}
\end{equation}
The preceding spectral bounds permit
$R\to\infty$. Using
Eqs.~\eqref{eq:caputo_diffusive_representation} and
\eqref{eq:integrated_fractional_power} yields
\begin{equation}
\mathcal{F}_{\alpha_j}(t)
=
\frac{d}{dt}
\mathcal{U}_{\alpha_j}^{\mathrm{diff}}(t)
+
\mathcal{Q}_{\alpha_j}^{\mathrm{diff}}(t)
\label{eq:integrated_diffusive_power_split}
\end{equation}
for almost every $t>0$. The variables
$\boldsymbol{\phi}_{j}(\xi,t)$ are auxiliary memory states that represent exponentially weighted moments of the same polarization history. They are not additional physical polarization channels.
To compare this realization with the history decomposition, write
\begin{subequations}
\label{eq:history_functional_identifications}
\begin{align}
\mathcal{U}_{\alpha_j}^{\mathrm{hist}}
&\equiv
\mathcal{U}_{\alpha_j},
\label{eq:history_storage_identification}
\\
\mathcal{Q}_{\alpha_j}^{\mathrm{hist}}
&\equiv
\mathcal{Q}_{\alpha_j}.
\label{eq:history_dissipation_identification}
\end{align}
\end{subequations}
where the right-hand sides are defined in
Eqs.~\eqref{eq:integrated_history_energy} and
\eqref{eq:integrated_history_dissipation}. For
$0\leq u\leq t$, define
\begin{equation}
\mathbf{h}_{j}(u,t)
=
\mathbf{P}_{j}(t)
-
\mathbf{P}_{j}(t-u).
\end{equation}
The quiescent initial condition gives
$\mathbf{h}_{j}(0,t)=\mathbf{0}$ and
$\mathbf{h}_{j}(t,t)=\mathbf{P}_{j}(t)$.
Differentiating the kernel representation with respect to $u$ gives
\begin{equation}
\frac{\alpha_j}{\Gamma(1-\alpha_j)}
u^{-\alpha_j-1}
=
\int_{0}^{\infty}
\xi e^{-\xi u}
\mu_{\alpha_j}(\xi)\,d\xi.
\end{equation}
Tonelli's theorem then transforms the history storage into
\begin{equation}
\begin{aligned}
\mathcal{U}_{\alpha_j}^{\mathrm{hist}}(t)
&=
\frac{\tau_j^{\alpha_j}}
{2\epsilon_0}
\int_{0}^{\infty}
\mu_{\alpha_j}(\xi)
\Bigg[
e^{-\xi t}
\left\|
\mathbf{P}_{j}(t)
\right\|_{j,\Omega}^{2}
\\
&\qquad\qquad+
\int_{0}^{t}
\xi e^{-\xi u}
\left\|
\mathbf{h}_{j}(u,t)
\right\|_{j,\Omega}^{2}\,du
\Bigg]d\xi.
\end{aligned}
\label{eq:integrated_history_energy_spectral_form}
\end{equation}
Changing variables in
Eq.~\eqref{eq:diffusive_variable}
and integrating by parts in
$L^{2}(\Omega)^{3}$ gives
\begin{equation}
\boldsymbol{\phi}_{j}(\xi,t)
=
e^{-\xi t}\mathbf{P}_{j}(t)
+
\int_{0}^{t}
\xi e^{-\xi u}
\mathbf{h}_{j}(u,t)\,du.
\label{eq:diffusive_variable_history_form}
\end{equation}
The coefficients form a probability measure because
\begin{equation}
e^{-\xi t}
+
\int_{0}^{t}
\xi e^{-\xi u}\,du
=
1.
\end{equation}
Strict convexity of the squared Hilbert norm and Jensen's inequality therefore imply
\begin{equation}
\left\|
\boldsymbol{\phi}_{j}(\xi,t)
\right\|_{j,\Omega}^{2}
\leq
e^{-\xi t}
\left\|
\mathbf{P}_{j}(t)
\right\|_{j,\Omega}^{2}
+
\int_{0}^{t}
\xi e^{-\xi u}
\left\|
\mathbf{h}_{j}(u,t)
\right\|_{j,\Omega}^{2}\,du.
\end{equation}
After multiplication by the positive spectral density and integration over $\xi$,
\begin{equation}
0
\leq
\mathcal{U}_{\alpha_j}^{\mathrm{diff}}(t)
\leq
\mathcal{U}_{\alpha_j}^{\mathrm{hist}}(t).
\label{eq:integrated_history_diffusive_energy_bound}
\end{equation}
The storage gap has an exact variance representation. Define
\begin{equation}
\begin{aligned}
\mathcal{V}_{j}(\xi,t)
={}&
e^{-\xi t}
\left\|
\mathbf{P}_{j}(t)
-
\boldsymbol{\phi}_{j}(\xi,t)
\right\|_{j,\Omega}^{2}
\\
&+
\int_{0}^{t}
\xi e^{-\xi u}
\left\|
\mathbf{h}_{j}(u,t)
-
\boldsymbol{\phi}_{j}(\xi,t)
\right\|_{j,\Omega}^{2}\,du.
\end{aligned}
\end{equation}
The Hilbert-space variance identity gives
\begin{equation}
\mathcal{U}_{\alpha_j}^{\mathrm{hist}}(t)
-
\mathcal{U}_{\alpha_j}^{\mathrm{diff}}(t)
=
\frac{\tau_j^{\alpha_j}}
{2\epsilon_0}
\int_{0}^{\infty}
\mu_{\alpha_j}(\xi)
\mathcal{V}_{j}(\xi,t)\,d\xi.
\label{eq:exact_memory_energy_gap}
\end{equation}
Thus, the difference between the two admissible storages is the positive weighted variance of the polarization increments around their diffusive mean.
The equality condition follows from strict convexity. Equality in
Eq.~\eqref{eq:integrated_history_diffusive_energy_bound}
requires
$\mathcal{V}_{j}(\xi,t)=0$
for $\mu_{\alpha_j}$-almost every $\xi$. For such $\xi$,
\begin{equation}
\mathbf{P}_{j}(t)
=
\mathbf{h}_{j}(u,t)
=
\boldsymbol{\phi}_{j}(\xi,t)
\end{equation}
for almost every $u\in(0,t)$. Since
$\mathbf{h}_{j}(\cdot,t)$ is continuous and
$\mathbf{h}_{j}(0,t)=\mathbf{0}$, this is possible only when
$\mathbf{P}_{j}(s)=\mathbf{0}$ for every $s\in[0,t]$. Conversely, a zero polarization history makes both storages vanish. Therefore,
\begin{equation}
\mathcal{U}_{\alpha_j}^{\mathrm{diff}}(t)
=
\mathcal{U}_{\alpha_j}^{\mathrm{hist}}(t)
\quad
\Longleftrightarrow
\quad
\mathbf{P}_{j}(s)=\mathbf{0}
\ \text{for all }s\in[0,t].
\label{eq:exact_energy_equality_condition}
\end{equation}
Every nontrivial continuous quiescent history satisfies the strict ordering
\begin{equation}
0
\leq
\mathcal{U}_{\alpha_j}^{\mathrm{diff}}(t)
<
\mathcal{U}_{\alpha_j}^{\mathrm{hist}}(t).
\end{equation}
Since the history storage vanishes at initialization,
Eq.~\eqref{eq:integrated_history_diffusive_energy_bound}
also gives
\begin{equation}
\lim_{t\to0^{+}}
\mathcal{U}_{\alpha_j}^{\mathrm{diff}}(t)
= 0.
\end{equation}
Integrating
Eq.~\eqref{eq:integrated_diffusive_power_split}
over $[\delta,t]$ and taking
$\delta\to0^{+}$ yields
\begin{equation}
\int_{0}^{t}
\mathcal{F}_{\alpha_j}(\eta)\,d\eta
=
\mathcal{U}_{\alpha_j}^{\mathrm{diff}}(t)
+
\int_{0}^{t}
\mathcal{Q}_{\alpha_j}^{\mathrm{diff}}(\eta)\,d\eta.
\label{eq:integrated_diffusive_energy_identity}
\end{equation}
The left-hand side is finite because
$\mathcal{F}_{\alpha_j}\in L^{1}(0,T)$, as established in
Sec.~\ref{sec:energy_decomposition}. Hence
$\mathcal{Q}_{\alpha_j}^{\mathrm{diff}}\in L^{1}(0,T)$.
As in the history representation, the instantaneous dissipation may contain an integrable fractional singularity at $t=0$.
The two decompositions reproduce the same constitutive power.
\begin{equation}
\mathcal{F}_{\alpha_j}
=
\frac{d\mathcal{U}_{\alpha_j}^{\mathrm{hist}}}{dt}
+
\mathcal{Q}_{\alpha_j}^{\mathrm{hist}}
=
\frac{d\mathcal{U}_{\alpha_j}^{\mathrm{diff}}}{dt}
+
\mathcal{Q}_{\alpha_j}^{\mathrm{diff}}.
\label{eq:two_memory_power_decompositions}
\end{equation}
Define
\begin{equation}
\Delta\mathcal{U}_{\alpha_j}(t)
=
\mathcal{U}_{\alpha_j}^{\mathrm{hist}}(t)
-
\mathcal{U}_{\alpha_j}^{\mathrm{diff}}(t)
\geq0.
\end{equation}
Subtracting the two power identities gives
\begin{equation}
\frac{d}{dt}
\Delta\mathcal{U}_{\alpha_j}(t)
=
\mathcal{Q}_{\alpha_j}^{\mathrm{diff}}(t)
-
\mathcal{Q}_{\alpha_j}^{\mathrm{hist}}(t).
\end{equation}
Both storages vanish at initialization, and therefore
\begin{equation}
\Delta\mathcal{U}_{\alpha_j}(t)
=
\int_{0}^{t}
\left[
\mathcal{Q}_{\alpha_j}^{\mathrm{diff}}(\eta)
-
\mathcal{Q}_{\alpha_j}^{\mathrm{hist}}(\eta)
\right]d\eta.
\label{eq:integrated_memory_storage_difference}
\end{equation}
Since the left-hand side is non-negative,
\begin{equation}
\int_{0}^{t}
\mathcal{Q}_{\alpha_j}^{\mathrm{diff}}(\eta)\,d\eta
\geq
\int_{0}^{t}
\mathcal{Q}_{\alpha_j}^{\mathrm{hist}}(\eta)\,d\eta.
\label{eq:accumulated_dissipation_ordering}
\end{equation}
This is an ordering of accumulated dissipation only. It does not imply a pointwise ordering between the two instantaneous rates, because
$d\Delta\mathcal{U}_{\alpha_j}/dt$
may change sign.
The comparison has a limited physical meaning. The history functional retains the full quadratic spread of the increments
$\mathbf{P}_{j}(t)-\mathbf{P}_{j}(s)$,
whereas the diffusive functional stores their exponentially weighted moments. The variance in
Eq.~\eqref{eq:exact_memory_energy_gap}
measures the quadratic information removed by this moment representation. It is not an additional physical energy reservoir. Both functionals satisfy the same constitutive power balance and are admissible storages for the same hereditary material law. Their ordering does not establish that either functional is a unique, minimum, or maximum thermodynamic free energy. Materials with memory can admit several compatible storage functionals \cite{Fabrizio:2014,AmendolaFabrizioGolden2021}.
If the stronger pointwise-in-space regularity used for
Eqs.~\eqref{eq:history_energy_functional} and
\eqref{eq:history_dissipation_functional}
holds, define
\begin{subequations}
\label{eq:local_diffusive_functionals}
\begin{align}
U_{\alpha_j}^{\mathrm{diff}}(\mathbf{r},t)
&=
\frac{\tau_j^{\alpha_j}}
{2\epsilon_0}
\int_{0}^{\infty}
\mu_{\alpha_j}(\xi)
\left\|
\boldsymbol{\phi}_{j}(\mathbf{r},\xi,t)
\right\|_{\boldsymbol{\chi}_{j}^{-1}}^{2}\,d\xi,
\label{eq:local_diffusive_storage}
\\
Q_{\alpha_j}^{\mathrm{diff}}(\mathbf{r},t)
&=
\frac{\tau_j^{\alpha_j}}
{\epsilon_0}
\int_{0}^{\infty}
\xi\mu_{\alpha_j}(\xi)
\left\|
\boldsymbol{\phi}_{j}(\mathbf{r},\xi,t)
\right\|_{\boldsymbol{\chi}_{j}^{-1}}^{2}\,d\xi.
\label{eq:local_diffusive_dissipation}
\end{align}
\end{subequations}
They satisfy
\begin{subequations}
\label{eq:local_diffusive_power_and_ordering}
\begin{align}
F_{\alpha_j}
&=
\partial_t U_{\alpha_j}^{\mathrm{diff}}
+
Q_{\alpha_j}^{\mathrm{diff}},
\label{eq:local_diffusive_power_split}
\\
0
\leq
U_{\alpha_j}^{\mathrm{diff}}
&\leq
U_{\alpha_j}^{\mathrm{hist}}.
\label{eq:local_diffusive_storage_ordering}
\end{align}
\end{subequations}
for almost every
$(\mathbf{r},t)\in\Omega\times(0,T)$.
Their spatial integrals reproduce the Hilbert-space quantities.
\begin{subequations}
\label{eq:spatial_diffusive_functionals}
\begin{align}
\mathcal{U}_{\alpha_j}^{\mathrm{diff}}(t)
&=
\int_{\Omega}
U_{\alpha_j}^{\mathrm{diff}}(\mathbf{r},t)\,dV,
\label{eq:spatial_diffusive_storage}
\\
\mathcal{Q}_{\alpha_j}^{\mathrm{diff}}(t)
&=
\int_{\Omega}
Q_{\alpha_j}^{\mathrm{diff}}(\mathbf{r},t)\,dV.
\label{eq:spatial_diffusive_dissipation}
\end{align}
\end{subequations}
The continuum representation also generates passive finite-dimensional approximations. Following standard positive-quadrature approximations of diffusive representations \cite{YuanAgrawal2002,Diethelm2022}, replace the spectral measure by
\begin{subequations}
\label{eq:positive_diffusive_quadrature}
\begin{align}
\mu_{\alpha_j}(\xi)\,d\xi
&\approx
\sum_{m=1}^{M}
w_{jm}
\delta(\xi-\xi_{jm}),
\label{eq:diffusive_quadrature_measure}
\\
w_{jm}&>0,
\label{eq:positive_diffusive_quadrature_weights}
\\
\xi_{jm}&>0.
\label{eq:positive_diffusive_quadrature_nodes}
\end{align}
\end{subequations}
The discrete states satisfy
\begin{equation}
\partial_t\boldsymbol{\phi}_{jm}
=
-\xi_{jm}\boldsymbol{\phi}_{jm}
+
\partial_t\mathbf{P}_{j},
\end{equation}
and their storage and dissipation are
\begin{subequations}
\label{eq:finite_diffusive_energy_dissipation}
\begin{align}
\mathcal{U}_{\alpha_j,M}^{\mathrm{diff}}
&=
\frac{\tau_j^{\alpha_j}}
{2\epsilon_0}
\sum_{m=1}^{M}
w_{jm}
\left\|
\boldsymbol{\phi}_{jm}
\right\|_{j,\Omega}^{2},
\label{eq:finite_diffusive_energy}
\\
\mathcal{Q}_{\alpha_j,M}^{\mathrm{diff}}
&=
\frac{\tau_j^{\alpha_j}}
{\epsilon_0}
\sum_{m=1}^{M}
\xi_{jm}w_{jm}
\left\|
\boldsymbol{\phi}_{jm}
\right\|_{j,\Omega}^{2}.
\label{eq:finite_diffusive_dissipation}
\end{align}
\end{subequations}
Positive nodes and weights preserve the non-negativity of the discrete storage and dissipation and the algebraic passivity structure of the continuum realization \cite{BaranowskiBauerMularczyk2021,Diethelm2022,ChaudharyDiethelm2024}. Positivity alone does not prove that the quadrature accurately approximates the Cole--Cole kernel or that a fully discrete Maxwell scheme converges. Spectral truncation, quadrature error, temporal discretization, and coupled-field convergence require separate analysis.

\section{Wave propagation and consistency checks}
\label{sec:wave_validation}
The finite-terminal constitutive law in
Eq.~\eqref{eq:cole_cole_constitutive}
contains both the transient produced by initialization at $t=0$ and the stationary response to periodic forcing. These contributions must be separated before a frequency-dependent permittivity is introduced. The local constitutive problem is first solved for a prescribed harmonic field. Its long-time limit then gives the source-free Maxwell eigenproblem. The prescribed field alone is not a finite-time solution of the coupled source-free Maxwell system.
Let an external source impose
\begin{equation}
\mathbf{E}_{\mathrm{test}}(\mathbf{r},t)
=
\mathbf{E}_{\mathrm{a}}
e^{i\mathbf{k}\cdot\mathbf{r}-i\omega t}
H(t),
\qquad
\omega>0,
\label{eq:harmonic_electric_field}
\end{equation}
where $\mathbf{E}_{\mathrm{a}}\in\mathbb{C}^{3}$ is the complex electric-field amplitude, $\mathbf{k}$ is a prescribed wave vector, $\omega>0$ is the angular frequency, $i^{2}=-1$ is the imaginary unit, and $H(t)$ is the Heaviside switching function. The physical field is the real part of
$\mathbf{E}_{\mathrm{test}}$. Abrupt switching gives
$\mathbf{E}_{\mathrm{test}}(\mathbf{r},0^{+})
=
\mathbf{E}_{\mathrm{a}}e^{i\mathbf{k}\cdot\mathbf{r}}$,
whereas
$\mathbf{P}_{j}(\mathbf{r},0)=\mathbf{0}$.
The resulting fractional initial layer is permitted by
Eq.~\eqref{eq:weighted_fractional_class}. A finite-time Maxwell realization would require a compatible imposed current, boundary source, or incident field. Until the stationary eigenproblem is introduced, the spatial phase in
Eq.~\eqref{eq:harmonic_electric_field}
is only a parameter of the local constitutive test.
Suppressing the common spatial factor, let
$\widehat{\mathbf{P}}_{j}(s)$ denote the Laplace transform of the polarization amplitude. Here $s$ is the complex Laplace variable. For $0<\alpha_j<1$, the standard Laplace transform of the Caputo derivative is \cite{Mainardi2022,GarrappaMainardiMaione2016}
\begin{equation}
\mathcal{L}
\left\{
{}^{C}_{0}D_t^{\alpha_j}\mathbf{P}_{j}
\right\}
=
s^{\alpha_j}\widehat{\mathbf{P}}_{j}(s)
-
s^{\alpha_j-1}\mathbf{P}_{j}(0).
\end{equation}
Using the quiescent initial condition and
$\mathcal{L}\{e^{-i\omega t}\}=(s+i\omega)^{-1}$,
Eq.~\eqref{eq:cole_cole_constitutive} gives
\begin{equation}
\widehat{\mathbf{P}}_{j}(s)
=
\frac{
\epsilon_0
\boldsymbol{\chi}_{j}
\mathbf{E}_{\mathrm{a}}}
{
(s+i\omega)
\left(
1+\tau_j^{\alpha_j}s^{\alpha_j}
\right)}.
\label{eq:polarization_laplace_solution}
\end{equation}
Using the corresponding Mittag--Leffler transform pair \cite{Mainardi2022,GarrappaMainardiMaione2016}, inversion gives
\begin{equation}
\begin{aligned}
\mathbf{P}_{j}(t)
&=
\frac{\epsilon_0}{\tau_j^{\alpha_j}}
\int_{0}^{t}
(t-u)^{\alpha_j-1}
E_{\alpha_j,\alpha_j}
\left[
-\left(
\frac{t-u}{\tau_j}
\right)^{\alpha_j}
\right]
\\
&\quad\times
\boldsymbol{\chi}_{j}
\mathbf{E}_{\mathrm{a}}
e^{-i\omega u}\,du,
\end{aligned}
\label{eq:finite_time_harmonic_polarization}
\end{equation}
where
$E_{a,b}(z)=
\sum_{n=0}^{\infty}z^{n}/\Gamma(an+b)$
is the two-parameter Mittag--Leffler function.
Equation~\eqref{eq:finite_time_harmonic_polarization}
solves the original finite-terminal Caputo problem. The lower terminal need not be replaced by $-\infty$.
Its short-time expansion is
\begin{subequations}
\label{eq:harmonic_initial_layer}
\begin{align}
\mathbf{P}_{j}(t)
&=
\frac{
\epsilon_0
\boldsymbol{\chi}_{j}
\mathbf{E}_{\mathrm{a}}}
{
\tau_j^{\alpha_j}
\Gamma(1+\alpha_j)}
t^{\alpha_j}
+
o(t^{\alpha_j}),
\label{eq:harmonic_initial_polarization}
\\
\partial_t\mathbf{P}_{j}(t)
&=
\frac{
\epsilon_0
\boldsymbol{\chi}_{j}
\mathbf{E}_{\mathrm{a}}}
{
\tau_j^{\alpha_j}
\Gamma(\alpha_j)}
t^{\alpha_j-1}
+
o(t^{\alpha_j-1}).
\label{eq:harmonic_initial_polarization_rate}
\end{align}
\end{subequations}
The abruptly switched harmonic forcing therefore reproduces the integrable initial singularity derived in
Sec.~\ref{sec:fractional_model}. For nonzero
$\mathbf{E}_{\mathrm{a}}$,
the polarization is generally not in
$C^{1}([0,T];L^{2}(\Omega)^{3})$.
To isolate the stationary response, define
\begin{equation}
g_{\alpha_j}(v)
=
\frac{v^{\alpha_j-1}}{\tau_j^{\alpha_j}}
E_{\alpha_j,\alpha_j}
\left[
-\left(
\frac{v}{\tau_j}
\right)^{\alpha_j}
\right].
\end{equation}
Changing variables in
Eq.~\eqref{eq:finite_time_harmonic_polarization}
gives
\begin{equation}
\mathbf{P}_{j}(t)
=
\epsilon_0
\boldsymbol{\chi}_{j}
\mathbf{E}_{\mathrm{a}}
e^{-i\omega t}
\int_{0}^{t}
g_{\alpha_j}(v)e^{i\omega v}\,dv.
\end{equation}
Since
\begin{equation}
\widehat{g}_{\alpha_j}(s)
=
\frac{1}
{1+\tau_j^{\alpha_j}s^{\alpha_j}},
\end{equation}
continuation from $\operatorname{Re}s>0$ to the causal boundary value $s=-i\omega+0^{+}$ gives \cite{StefanskiGulgowskiTsakmakidis2022}
\begin{equation}
\int_{0}^{\infty}
g_{\alpha_j}(v)e^{i\omega v}\,dv
=
\frac{1}
{1+(-i\omega\tau_j)^{\alpha_j}}.
\label{eq:harmonic_kernel_fourier_limit}
\end{equation}
The large-argument Mittag--Leffler expansion gives
$g_{\alpha_j}(v)=\mathcal{O}(v^{-\alpha_j-1})$,
so the absolute tail is
$\int_t^\infty|g_{\alpha_j}(v)|\,dv
=
\mathcal{O}(t^{-\alpha_j})$
\cite{Mainardi2022,GarrappaMainardiMaione2016}. Hence
\begin{subequations}
\label{eq:stationary_polarization_limit}
\begin{align}
\mathbf{P}_{j}(t)
&=
\widetilde{\mathbf{P}}_{j}(\omega)e^{-i\omega t}
+
\mathcal{O}(t^{-\alpha_j}),
\qquad
t\to\infty,
\label{eq:polarization_long_time}
\\
\widetilde{\mathbf{P}}_{j}(\omega)
&=
\frac{
\epsilon_0\boldsymbol{\chi}_{j}}
{
1+(-i\omega\tau_j)^{\alpha_j}}
\mathbf{E}_{\mathrm{a}}.
\label{eq:steady_polarization_amplitude}
\end{align}
\end{subequations}
The causal fractional power is evaluated on the branch
\begin{equation}
(-i\omega\tau_j)^{\alpha_j}
=
(\omega\tau_j)^{\alpha_j}
e^{-i\pi\alpha_j/2}.
\label{eq:fractional_frequency_branch}
\end{equation}
Thus, the stationary Cole--Cole susceptibility is the long-time boundary value of the finite-terminal initial-value problem rather than an independent constitutive assumption. The causal boundary value is consistent with the analytic structure required of electromagnetic response functions \cite{StefanskiGulgowskiTsakmakidis2022}.
Assume now that
$\boldsymbol{\epsilon}_{\infty}$,
$\boldsymbol{\chi}_{j}$, and
$\boldsymbol{\sigma}$
are spatially constant, that the medium is unbounded or locally homogeneous on the propagation scale, and that
$\mathbf{J}_{\mathrm{ext}}=\mathbf{0}$
in the stationary regime. For the convention
$e^{-i\omega t}$, extending the established scalar Cole--Cole response \cite{ColeCole1941,GarrappaMainardiMaione2016} to $N$ anisotropic channels and including the separated conductive current gives
\begin{equation}
\widetilde{\boldsymbol{\epsilon}}_{\mathrm{r}}(\omega)
=
\boldsymbol{\epsilon}_{\infty}
+
\sum_{j=1}^{N}
\frac{
\boldsymbol{\chi}_{j}}
{
1+(-i\omega\tau_j)^{\alpha_j}}
+
\frac{
i\boldsymbol{\sigma}}
{\epsilon_0\omega}.
\label{eq:generalized_cole_cole_permittivity}
\end{equation}
The first term is the high-frequency displacement, the sum describes the relaxing polarization channels, and the final term incorporates conduction into the effective displacement.
For a complex tensor $\mathbf{A}$, let
$\operatorname{Im}_{\mathrm{H}}\mathbf{A}
=
(\mathbf{A}-\mathbf{A}^{\dagger})/(2i)$.
Set
\begin{subequations}
\label{eq:cole_cole_frequency_variables}
\begin{align}
x_j&=(\omega\tau_j)^{\alpha_j},
\label{eq:cole_cole_frequency_magnitude}
\\
\theta_j&=\frac{\pi\alpha_j}{2},
\label{eq:cole_cole_frequency_angle}
\\
\Delta_j
&=
1+2x_j\cos\theta_j+x_j^{2}.
\label{eq:cole_cole_frequency_denominator}
\end{align}
\end{subequations}
Because all material tensors are real and symmetric, for every
$\mathbf{z}\in\mathbb{C}^{3}$,
\begin{equation}
\mathbf{z}^{\dagger}
\operatorname{Im}_{\mathrm{H}}
\widetilde{\boldsymbol{\epsilon}}_{\mathrm{r}}(\omega)
\mathbf{z}
=
\sum_{j=1}^{N}
\frac{x_j\sin\theta_j}{\Delta_j}
\mathbf{z}^{\dagger}
\boldsymbol{\chi}_{j}\mathbf{z}
+
\frac{
\mathbf{z}^{\dagger}
\boldsymbol{\sigma}\mathbf{z}}
{\epsilon_0\omega}
\geq 0.
\label{eq:frequency_passivity}
\end{equation}
The inequality follows from
$\boldsymbol{\chi}_{j}\succ 0$,
$\boldsymbol{\sigma}\succeq 0$, and
$0<\theta_j<\pi/2$. For the $e^{-i\omega t}$ convention, the standard cycle-averaged material absorption density is \cite{Jackson:1999}
\begin{equation}
\left\langle
p_{\mathrm{abs}}
\right\rangle
=
\frac{\omega\epsilon_0}{2}
\mathbf{E}_{\mathrm{a}}^{\dagger}
\operatorname{Im}_{\mathrm{H}}
\widetilde{\boldsymbol{\epsilon}}_{\mathrm{r}}(\omega)
\mathbf{E}_{\mathrm{a}}
\geq 0.
\label{eq:harmonic_absorbed_power}
\end{equation}
This is the stationary frequency-domain counterpart of the time-domain passivity inequality.
Let
\begin{subequations}
\label{eq:harmonic_field_amplitudes}
\begin{align}
\mathbf{E}_{\omega}(\mathbf{r})
&=
\mathbf{E}_{\mathrm{a}}e^{i\mathbf{k}\cdot\mathbf{r}},
\label{eq:harmonic_electric_amplitude}
\\
\mathbf{H}_{\omega}(\mathbf{r})
&=
\mathbf{H}_{\mathrm{a}}e^{i\mathbf{k}\cdot\mathbf{r}},
\label{eq:harmonic_magnetic_amplitude}
\end{align}
\end{subequations}
where $\mathbf{k}$ may be complex. Maxwell's curl equations and the divergence constraint give
\begin{subequations}
\label{eq:anisotropic_wave_system}
\begin{align}
\mathbf{k}\times
\left(
\mathbf{k}\times\mathbf{E}_{\mathrm{a}}
\right)
+
\frac{\omega^{2}}{c^{2}}
\widetilde{\boldsymbol{\epsilon}}_{\mathrm{r}}(\omega)
\mathbf{E}_{\mathrm{a}}
&=
\mathbf{0},
\label{eq:anisotropic_wave_eigenproblem}
\\
\mathbf{k}\cdot
\widetilde{\boldsymbol{\epsilon}}_{\mathrm{r}}(\omega)
\mathbf{E}_{\mathrm{a}}
&=
0,
\label{eq:anisotropic_transversality}
\end{align}
\end{subequations}
where
$c=(\mu_0\epsilon_0)^{-1/2}$.
These equations determine the admissible polarization vectors and complex wave numbers of a homogeneous anisotropic medium. Time-fractional electromagnetic models produce the same connection between temporal memory, dispersion, and wave attenuation in direct propagation analyses
\cite{LiHuangLin2011,PietruszkaStefanskiGulgowski2023}.
In an isotropic medium,
$\widetilde{\boldsymbol{\epsilon}}_{\mathrm{r}}
=
\widetilde{\epsilon}_{\mathrm{r}}\mathbf{I}$.
Provided
$\widetilde{\epsilon}_{\mathrm{r}}\neq 0$,
Eq.~\eqref{eq:anisotropic_transversality}
gives
$\mathbf{k}\cdot\mathbf{E}_{\mathrm{a}}=0$, while
Eq.~\eqref{eq:anisotropic_wave_eigenproblem}
reduces to
\begin{equation}
k^{2}
=
\frac{\omega^{2}}{c^{2}}
\widetilde{\epsilon}_{\mathrm{r}}(\omega).
\label{eq:isotropic_dispersion_relation}
\end{equation}
Writing
$\widetilde{\epsilon}_{\mathrm{r}}
=
\epsilon'+i\epsilon''$,
the real and imaginary parts are
\begin{subequations}
\label{eq:isotropic_permittivity_parts}
\begin{align}
\epsilon'(\omega)
&=
\epsilon_{\infty}
+
\sum_{j=1}^{N}
\chi_j
\frac{
1+x_j\cos\theta_j}
{\Delta_j},
\label{eq:real_permittivity}
\\
\epsilon''(\omega)
&=
\sum_{j=1}^{N}
\chi_j
\frac{
x_j\sin\theta_j}
{\Delta_j}
+
\frac{\sigma}{\epsilon_0\omega}.
\label{eq:imaginary_permittivity}
\end{align}
\end{subequations}
Here,
$\epsilon_{\infty}$, $\chi_j$, and $\sigma$
are the scalar isotropic material coefficients.
For propagation in a chosen direction, take the passive square-root branch
\begin{subequations}
\label{eq:passive_wave_number_branch}
\begin{align}
k(\omega)
&=
\beta(\omega)+i\gamma(\omega),
\label{eq:complex_wave_number_decomposition}
\\
\beta&\geq 0,
\label{eq:nonnegative_phase_constant}
\\
\gamma&\geq 0.
\label{eq:nonnegative_attenuation_constant}
\end{align}
\end{subequations}
The factor
$e^{ikz}=e^{i\beta z-\gamma z}$
then decays along the propagation direction. Separating the real and imaginary parts of
Eq.~\eqref{eq:isotropic_dispersion_relation}
gives
\begin{subequations}
\label{eq:propagation_constants}
\begin{align}
\beta(\omega)
&=
\frac{\omega}{c}
\left[
\frac{
|\widetilde{\epsilon}_{\mathrm{r}}|+\epsilon'}
{2}
\right]^{1/2},
\label{eq:phase_constant}
\\
\gamma(\omega)
&=
\frac{\omega}{c}
\left[
\frac{
|\widetilde{\epsilon}_{\mathrm{r}}|-\epsilon'}
{2}
\right]^{1/2},
\label{eq:attenuation_constant}
\end{align}
\end{subequations}
where
$|\widetilde{\epsilon}_{\mathrm{r}}|
=
[(\epsilon')^{2}+(\epsilon'')^{2}]^{1/2}$.
When $\gamma>0$, the amplitude penetration depth is
$\delta_{\mathrm{p}}=1/\gamma$. It is infinite for a lossless mode.
The stationary response reproduces the time-domain constitutive limits. At fixed $\omega$ and $\tau_j$,
\begin{subequations}
\label{eq:stationary_constitutive_limits}
\begin{align}
\frac{
\boldsymbol{\chi}_{j}}
{
1+(-i\omega\tau_j)^{\alpha_j}}
&\longrightarrow
\frac{
\boldsymbol{\chi}_{j}}
{
1-i\omega\tau_j},
&&
\alpha_j\to 1^{-},
\label{eq:debye_frequency_limit}
\\
\frac{
\boldsymbol{\chi}_{j}}
{
1+(-i\omega\tau_j)^{\alpha_j}}
&\longrightarrow
\boldsymbol{\chi}_{j},
&&
\tau_j\to 0^{+}.
\label{eq:instantaneous_frequency_limit}
\end{align}
\end{subequations}
The first limit is the Debye response, while the second makes the channel instantaneous.
When
$\boldsymbol{\sigma}=\mathbf{0}$,
\begin{equation}
\lim_{\omega\to 0^{+}}
\widetilde{\boldsymbol{\epsilon}}_{\mathrm{r}}(\omega)
=
\boldsymbol{\epsilon}_{\mathrm{s}}
=
\boldsymbol{\epsilon}_{\infty}
+
\sum_{j=1}^{N}
\boldsymbol{\chi}_{j}.
\label{eq:static_permittivity}
\end{equation}
For
$\boldsymbol{\sigma}\neq\mathbf{0}$,
only the polarization part has this limit. The conductive contribution
$i\boldsymbol{\sigma}/(\epsilon_0\omega)$
diverges as $\omega^{-1}$. Thus,
$\boldsymbol{\epsilon}_{\mathrm{s}}$
is the relaxed polarization tensor after conduction has been separated.
For
$\omega\tau_j\gg 1$,
\begin{equation}
\frac{
\boldsymbol{\chi}_{j}}
{
1+(-i\omega\tau_j)^{\alpha_j}}
=
\boldsymbol{\chi}_{j}
(\omega\tau_j)^{-\alpha_j}
e^{i\pi\alpha_j/2}
+
\mathcal{O}
\left(
(\omega\tau_j)^{-2\alpha_j}
\right).
\label{eq:high_frequency_channel_asymptotic}
\end{equation}
Each polarization channel therefore vanishes algebraically at high frequency. The conductive term also tends to zero as $\omega^{-1}$, so
$\widetilde{\boldsymbol{\epsilon}}_{\mathrm{r}}
\to
\boldsymbol{\epsilon}_{\infty}$.
Simple attenuation power laws require additional weak-loss assumptions. Consider one isotropic channel with
$\sigma=0$, a positive real background
$\epsilon_{\mathrm{b}}$, and
$\epsilon''_{j}\ll\epsilon_{\mathrm{b}}$. Expanding
Eq.~\eqref{eq:attenuation_constant}
gives
\begin{equation}
\gamma_j(\omega)
\simeq
\frac{\omega}
{2c\sqrt{\epsilon_{\mathrm{b}}}}
\epsilon''_{j}(\omega).
\label{eq:weak_loss_attenuation}
\end{equation}
For
$\omega\tau_j\ll 1$ and
$\omega\tau_j\gg 1$, respectively,
\begin{equation}
\epsilon''_{j}
=
\chi_j\sin\theta_j
(\omega\tau_j)^{\alpha_j}
+
\mathcal{O}
\left(
(\omega\tau_j)^{2\alpha_j}
\right)
\end{equation}
and
\begin{equation}
\epsilon''_{j}
=
\chi_j\sin\theta_j
(\omega\tau_j)^{-\alpha_j}
+
\mathcal{O}
\left(
(\omega\tau_j)^{-2\alpha_j}
\right).
\end{equation}
Consequently,
\begin{subequations}
\label{eq:attenuation_power_laws}
\begin{align}
\gamma_j(\omega)
&\propto
\omega^{1+\alpha_j},
&
\omega\tau_j
&\ll 1,
\label{eq:low_frequency_attenuation_scaling}
\\
\gamma_j(\omega)
&\propto
\omega^{1-\alpha_j},
&
\omega\tau_j
&\gg 1.
\label{eq:high_frequency_attenuation_scaling}
\end{align}
\end{subequations}
These relations require negligible conductivity, weak total loss, a slowly varying positive background, and no strongly overlapping relaxation bands. They are not general asymptotic laws for strongly lossy, conductive, anisotropic, or overlapping multichannel media.

The effect of the fractional order can be quantified for one isotropic, nonconductive channel. Let $u=\omega\tau$, $x=u^{\alpha}$, $\theta=\pi\alpha/2$, and define the channel-normalized dielectric loss $L_{\alpha}=\epsilon''/\chi$. Equations~\eqref{eq:real_permittivity} and \eqref{eq:imaginary_permittivity} give
\begin{subequations}
\label{eq:fractional_order_loss_metrics}
\begin{align}
L_{\alpha}(u)
&=
\frac{x\sin\theta}
{1+2x\cos\theta+x^{2}},
\label{eq:normalized_loss_function}
\\
\frac{dL_{\alpha}}{dx}
&=
\frac{(1-x^{2})\sin\theta}
{(1+2x\cos\theta+x^{2})^{2}},
\label{eq:normalized_loss_derivative}
\\
u_{\mathrm{peak}}
&=1,
\qquad
L_{\alpha}^{\max}
=
\frac{1}{2}
\tan\left(\frac{\pi\alpha}{4}\right),
\label{eq:normalized_loss_peak}
\\
x_{\pm}
&=
2+\cos\theta
\pm
\sqrt{(2+\cos\theta)^{2}-1},
\qquad
u_{\pm}=x_{\pm}^{1/\alpha},
\label{eq:loss_half_max_points}
\\
W_{\alpha}
&=
\log_{10}\left(\frac{u_{+}}{u_{-}}\right)
=
\frac{2}{\alpha}\log_{10}x_{+}.
\label{eq:loss_bandwidth}
\end{align}
\end{subequations}
Here, $u_{\mathrm{peak}}$ denotes the dimensionless loss-peak frequency, $L_{\alpha}^{\max}$ the normalized peak height, $x_{\pm}$ and $u_{\pm}$ the two half-maximum locations, and $W_{\alpha}$ the full width at half maximum in decades of $\omega\tau$. The last line uses $x_{+}x_{-}=1$. The loss peak remains at the characteristic frequency for every $\alpha$, but its height and logarithmic width change with $\alpha$. Table~\ref{tab:fractional_order_metrics} shows that reducing $\alpha$ from the Debye limit to $0.4$ decreases the normalized peak from $0.5000$ to $0.1625$ and broadens its half-maximum width from $1.1439$ to $3.6756$ decades.

\begin{table}[t]
\caption{Exact fractional-order metrics for a single Cole--Cole channel. The entry $\alpha=1$ denotes the Debye limit.}
\label{tab:fractional_order_metrics}
\centering
\begin{tabular}{ccc}
\toprule
$\alpha$ & $L_{\alpha}^{\max}$ & $W_{\alpha}$ (decades) \\
\midrule
$0.4$ & $0.162460$ & $3.675601$ \\
$0.6$ & $0.254763$ & $2.322512$ \\
$0.8$ & $0.363271$ & $1.606225$ \\
$1.0$ & $0.500000$ & $1.143895$ \\
\bottomrule
\end{tabular}
\end{table}

Table~\ref{tab:fractional_order_metrics} separates two effects of the order parameter. The characteristic frequency remains fixed at $u_{\mathrm{peak}}=1$. As $\alpha$ decreases, the peak height decreases and the half-maximum bandwidth increases. The table quantifies the spectral broadening shown by the loss curves below.

\begin{figure}[t]
\centering
\includegraphics[width=0.75\columnwidth]{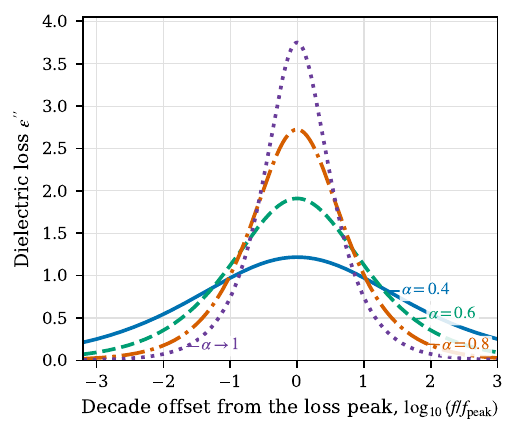}
\caption{Dielectric loss $\epsilon''$ for the illustrative single-channel medium with $\epsilon_{\infty}=2.5$, $\chi=7.5$, $\tau=1\,\mathrm{ns}$, and $\sigma=0$. Each curve is labeled inside the axes. The $\alpha\to1$ curve is the Debye limit. Decreasing $\alpha$ lowers and broadens the loss peak.}
\label{fig:fractional_order_loss}
\end{figure}

\begin{figure}[t]
\centering
\includegraphics[width=0.75\columnwidth]{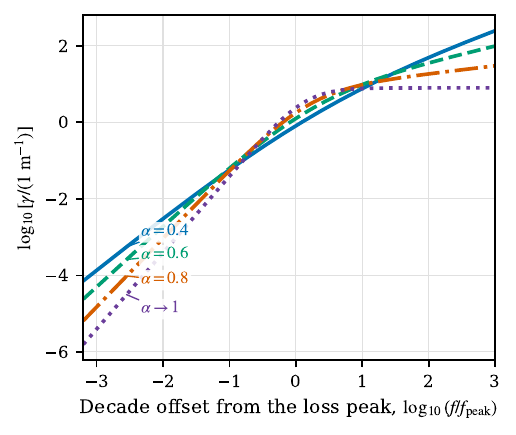}
\caption{Exact attenuation coefficient for the same single-channel medium, displayed as $\log_{10}[\gamma/(1\,\mathrm{m}^{-1})]$. Labels inside the axes identify $\alpha$. Decreasing $\alpha$ redistributes attenuation toward frequencies farther from $\omega\tau=1$, so the curves cross rather than remain globally ordered.}
\label{fig:fractional_order_attenuation}
\end{figure}

Figures~\ref{fig:fractional_order_loss} and \ref{fig:fractional_order_attenuation} give a dimensional example with $\epsilon_{\infty}=2.5$, $\chi=7.5$, $\tau=1\,\mathrm{ns}$, and $\sigma=0$. These illustrative values are not fitted to a specimen. They place the common loss-peak frequency at $f_{\mathrm{peak}}=(2\pi\tau)^{-1}=159.15\,\mathrm{MHz}$. The variable $u=\omega\tau$ controls the shape, while $\chi$ and $\tau$ set the vertical and horizontal scales. Both figures use the decade offset $\log_{10}(f/f_{\mathrm{peak}})=\log_{10}(\omega\tau)$. Zero marks the common loss peak, and each integer step represents one decade. The attenuation curves use the exact square-root expression in Eq.~\eqref{eq:attenuation_constant}, not the weak-loss approximation. Smaller $\alpha$ increases the low- and high-frequency loss tails but decreases the response near $u=1$. Thus, neither dielectric loss nor attenuation is globally monotone in $\alpha$. A smaller order spreads absorption and propagation loss over a broader band instead of uniformly increasing a loss coefficient.

Figure~\ref{fig:fractional_order_loss} shows that changing $\alpha$ redistributes dielectric loss around the fixed characteristic frequency rather than shifting the peak. The lower-order curves are flatter near the center and larger in the low- and high-frequency tails, as expected for a broader relaxation-time distribution.

Figure~\ref{fig:fractional_order_attenuation} shows the corresponding redistribution of wave attenuation through the exact complex square root of the permittivity. The curves cross because a smaller fractional order does not give uniformly larger attenuation. Their ordering depends on the observation frequency relative to $\omega\tau=1$.

The same scaling applies to fitted Cole--Cole channels. Changing $\tau$ shifts the response along the frequency axis, changing $\chi$ changes the polarization strength, and summing channels produces a multiband spectrum. Parametric tissue models are one application of such multichannel fits \cite{GabrielLauGabriel1996}. For these models, the energy result gives a passive interpretation of each channel. Figures~\ref{fig:fractional_order_loss} and \ref{fig:fractional_order_attenuation} also show why uncertainty in $\alpha$ affects broadband attenuation rather than the loss at only one frequency.

An exact transient check is obtained from the externally prescribed step
\begin{equation}
\mathbf{E}_{\mathrm{step}}(t)
=
\mathbf{E}_{\mathrm{s}}H(t),
\label{eq:step_electric_field}
\end{equation}
where $\mathbf{E}_{\mathrm{s}}$ is a constant real vector. This is again a local constitutive test rather than a source-free spatially uniform Maxwell wave. Applying the established scalar Cole--Cole step response \cite{GarrappaMainardiMaione2016,Mainardi2022} componentwise to each anisotropic channel gives, for quiescent polarization,
\begin{equation}
\mathbf{P}_{j}(t)
=
\epsilon_0
\boldsymbol{\chi}_{j}
\mathbf{E}_{\mathrm{s}}
\left[
1-
E_{\alpha_j}
\left(
-\left(
\frac{t}{\tau_j}
\right)^{\alpha_j}
\right)
\right],
\label{eq:step_polarization_solution}
\end{equation}
where
$E_{\alpha}=E_{\alpha,1}$.
The derivative and Caputo derivative are
\begin{subequations}
\label{eq:step_polarization_rates}
\begin{align}
\partial_t\mathbf{P}_{j}(t)
&=
\frac{
\epsilon_0
\boldsymbol{\chi}_{j}
\mathbf{E}_{\mathrm{s}}}
{\tau_j^{\alpha_j}}
t^{\alpha_j-1}
E_{\alpha_j,\alpha_j}
\left[
-\left(
\frac{t}{\tau_j}
\right)^{\alpha_j}
\right],
\label{eq:step_polarization_derivative}
\\
{}^{C}_{0}D_t^{\alpha_j}
\mathbf{P}_{j}(t)
&=
\frac{
\epsilon_0
\boldsymbol{\chi}_{j}
\mathbf{E}_{\mathrm{s}}}
{\tau_j^{\alpha_j}}
E_{\alpha_j}
\left[
-\left(
\frac{t}{\tau_j}
\right)^{\alpha_j}
\right].
\label{eq:step_caputo_derivative}
\end{align}
\end{subequations}
Near $t=0$,
\begin{subequations}
\label{eq:step_initial_layer}
\begin{align}
\mathbf{P}_{j}(t)
&=
\frac{
\epsilon_0
\boldsymbol{\chi}_{j}
\mathbf{E}_{\mathrm{s}}}
{
\tau_j^{\alpha_j}
\Gamma(1+\alpha_j)}
t^{\alpha_j}
+
\mathcal{O}(t^{2\alpha_j}),
\label{eq:step_initial_polarization}
\\
\partial_t\mathbf{P}_{j}(t)
&=
\frac{
\epsilon_0
\boldsymbol{\chi}_{j}
\mathbf{E}_{\mathrm{s}}}
{
\tau_j^{\alpha_j}
\Gamma(\alpha_j)}
t^{\alpha_j-1}
+
\mathcal{O}(t^{2\alpha_j-1}).
\label{eq:step_initial_polarization_rate}
\end{align}
\end{subequations}
This exact solution realizes the fractional initial layer assumed in
Secs.~\ref{sec:fractional_model}--\ref{sec:global_energy}.
For $0<\alpha_j<1$, $E_{\alpha_j}(-x)$ is completely monotone and $t^{\alpha_j-1}E_{\alpha_j,\alpha_j}(-t^{\alpha_j})$ is non-negative \cite{Pollard1948,GarrappaMainardiMaione2016}. Substituting Eqs.~\eqref{eq:step_polarization_derivative} and \eqref{eq:step_caputo_derivative} into Eq.~\eqref{eq:fractional_power_density} therefore gives
\begin{equation}
\begin{aligned}
F_{\alpha_j}^{\mathrm{step}}(t)
&=
\frac{
\epsilon_0
\mathbf{E}_{\mathrm{s}}\cdot
\boldsymbol{\chi}_{j}
\mathbf{E}_{\mathrm{s}}}
{\tau_j^{\alpha_j}}
t^{\alpha_j-1}
E_{\alpha_j,\alpha_j}
\left[
-\left(
\frac{t}{\tau_j}
\right)^{\alpha_j}
\right]
\\
&\quad\times
E_{\alpha_j}
\left[
-\left(
\frac{t}{\tau_j}
\right)^{\alpha_j}
\right]
\geq 0.
\end{aligned}
\label{eq:step_fractional_power}
\end{equation}
Positive definiteness of $\boldsymbol{\chi}_{j}$ gives the stated inequality. Its initial form is
\begin{equation}
F_{\alpha_j}^{\mathrm{step}}(t)
=
\frac{
\epsilon_0
\mathbf{E}_{\mathrm{s}}\cdot
\boldsymbol{\chi}_{j}
\mathbf{E}_{\mathrm{s}}}
{
\tau_j^{\alpha_j}
\Gamma(\alpha_j)}
t^{\alpha_j-1}
+
o(t^{\alpha_j-1}),
\label{eq:step_power_initial_asymptotic}
\end{equation}
which may diverge at initialization but remains locally integrable.

For the isotropic single-channel step, set $y=t/\tau$, $R_{\alpha}(y)=E_{\alpha}(-y^{\alpha})$, and let $E_{\mathrm{s}}$ denote the scalar electric-field amplitude along the polarization direction. In the equations below, $P(t)$ is the scalar polarization component along that direction, $F_{\alpha}^{\mathrm{step}}(t)$ is the corresponding fractional polarization power density, $p_{\alpha}$ is the polarization normalized by its static channel amplitude, and $\Pi_{\alpha}$ is the dimensionless fractional power. The exact normalized observables are
\begin{subequations}
\label{eq:normalized_step_observables}
\begin{align}
p_{\alpha}(y)
&=
\frac{P(t)}{\epsilon_0\chi E_{\mathrm{s}}}
=
1-R_{\alpha}(y),
\label{eq:normalized_step_polarization}
\\
\Pi_{\alpha}(y)
&=
\frac{\tau F_{\alpha}^{\mathrm{step}}(t)}
{\epsilon_0\chi E_{\mathrm{s}}^{2}}
=
y^{\alpha-1}
E_{\alpha,\alpha}(-y^{\alpha})
R_{\alpha}(y).
\label{eq:normalized_step_power}
\end{align}
\end{subequations}
In the Debye limit these functions reduce to $p_1(y)=1-e^{-y}$ and $\Pi_1(y)=e^{-2y}$. For $0<\alpha<1$, the standard small-argument Mittag--Leffler asymptotics \cite{Mainardi2022,GarrappaMainardiMaione2016} give
\begin{subequations}
\label{eq:step_short_time_asymptotics}
\begin{align}
p_{\alpha}(y)
&\sim
\frac{y^{\alpha}}{\Gamma(1+\alpha)},
\qquad y\to0^{+},
\label{eq:step_polarization_short_time_asymptotic}
\\
\Pi_{\alpha}(y)
&\sim
\frac{y^{\alpha-1}}{\Gamma(\alpha)},
\qquad y\to0^{+}.
\label{eq:step_power_short_time_asymptotic}
\end{align}
\end{subequations}
The standard large-argument Mittag--Leffler asymptotics \cite{Mainardi2022,GarrappaMainardiMaione2016} give
\begin{subequations}
\label{eq:step_long_time_asymptotics}
\begin{align}
1-p_{\alpha}(y)
&\sim
\frac{y^{-\alpha}}{\Gamma(1-\alpha)},
\qquad y\to\infty,
\label{eq:step_polarization_long_time_asymptotic}
\\
\Pi_{\alpha}(y)
&\sim
\frac{\alpha y^{-2\alpha-1}}
{\Gamma(1-\alpha)^{2}},
\qquad y\to\infty.
\label{eq:step_power_long_time_asymptotic}
\end{align}
\end{subequations}
Lowering $\alpha$ has two distinct effects. It makes the initial power more singular and replaces Debye exponential settling with a slower algebraic tail. The initial singularity does not produce infinite energy because its time integral remains finite, as shown by Eq.~\eqref{eq:initial_accumulated_dissipation}.

\begin{figure}[t]
\centering
\includegraphics[width=0.75\columnwidth]{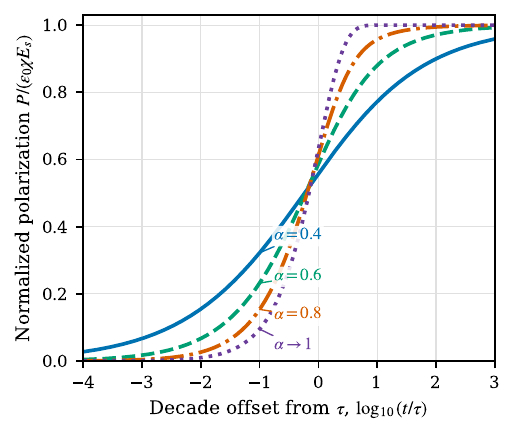}
\caption{Normalized polarization $p_{\alpha}=P/(\epsilon_{0}\chi E_{s})$ for the exact electric-step response. Labels inside the axes identify $\alpha$. A smaller $\alpha$ produces a slower long-time approach to equilibrium. The limit $\alpha\to1$ gives the Debye response.}
\label{fig:fractional_order_polarization}
\end{figure}

\begin{figure}[t]
\centering
\includegraphics[width=0.75\columnwidth]{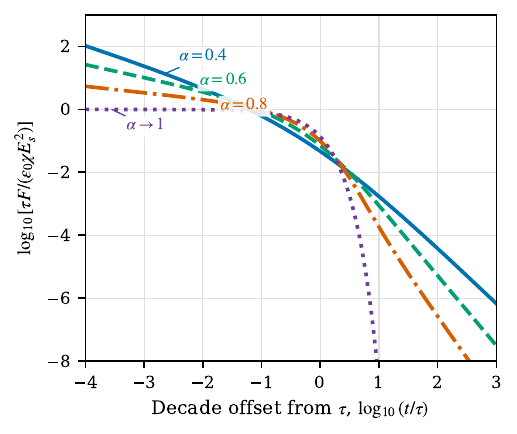}
\caption{Scaled fractional power from the exact electric-step response, displayed as $\log_{10}[\tau F/(\epsilon_{0}\chi E_{s}^{2})]$. Labels inside the axes identify $\alpha$. A smaller $\alpha$ gives a stronger integrable initial power layer and a slower algebraic tail. The limit $\alpha\to1$ gives the Debye response.}
\label{fig:fractional_order_power}
\end{figure}

The curves in Figs.~\ref{fig:fractional_order_polarization} and \ref{fig:fractional_order_power} were evaluated from the established positive relaxation-spectrum identity \cite{Pollard1948,GarrappaMainardiMaione2016}
\begin{equation}
R_{\alpha}(y)
=
\int_{0}^{\infty}
e^{-r y}
\frac{
r^{\alpha-1}\sin(\pi\alpha)}
{\pi\left[r^{2\alpha}+2r^{\alpha}\cos(\pi\alpha)+1\right]}
\,dr,
\qquad
0<\alpha<1,
\label{eq:mittag_leffler_positive_spectrum}
\end{equation}
and $-R_{\alpha}'(y)=y^{\alpha-1}E_{\alpha,\alpha}(-y^{\alpha})$. With $r=e^{z}$, the accompanying notebook applies the composite trapezoidal rule on $-55\leq z\leq35$ with $45001$ nodes and uses $620$ logarithmically spaced values of $10^{-4}\leq y\leq10^{3}$. At $y=0.01$, $1$, and $100$ for $\alpha=0.4$, $0.6$, and $0.8$, the maximum relative discrepancy from independent Wolfram Language evaluations of $R_{\alpha}$ and $-R_{\alpha}'$ is $3.13\times10^{-4}$. This comparison checks the numerical evaluation of the algebraic tail over the stated range. The horizontal coordinate in both plots is $\log_{10}(t/\tau)$. Zero marks the characteristic time, and each integer step represents one decade without scientific-notation tick labels.

Figure~\ref{fig:fractional_order_polarization} compares the approach to the static polarization level on a decade time scale. At late times, the lower-order responses remain farther from equilibrium and therefore retain longer memory.

Figure~\ref{fig:fractional_order_power} displays both temporal regimes of the fractional power. Near initialization, the lower-order curves rise because the integrable $t^{\alpha-1}$ layer is stronger. At late times, they remain above the Debye curve because the decay is algebraic rather than exponential. The same fractional order thus controls both the initial singularity and the long-time dissipation tail.

The step result shows the time-domain effect of the broadened frequency response. A smaller $\alpha$ represents a wider distribution of effective relaxation rates. Fast modes produce the sharp initial response, while slow modes retain polarization after the nominal time $\tau$. In pulse propagation or time-domain dielectric spectroscopy, this response gives longer post-excitation settling and extends the interval of material absorption. The example quantifies these effects without assigning a microscopic mechanism that the phenomenological Cole--Cole law does not determine.

For a homogeneous specimen of volume $|\Omega|$, let
$\mathcal{F}_{\alpha_j}^{\mathrm{step}}
=
|\Omega|F_{\alpha_j}^{\mathrm{step}}$
and define the supplied fractional work by
\begin{equation}
\mathcal{W}_{j}^{\mathrm{step}}(t_{\mathrm{f}})
=
\int_{0}^{t_{\mathrm{f}}}
\mathcal{F}_{\alpha_j}^{\mathrm{step}}(t)\,dt,
\qquad
t_{\mathrm{f}}>0.
\label{eq:step_fractional_work}
\end{equation}
For
$\mathbf{E}_{\mathrm{s}}\neq\mathbf{0}$,
Eq.~\eqref{eq:step_fractional_power}
and the positive definiteness of
$\boldsymbol{\chi}_{j}$
give
$\mathcal{W}_{j}^{\mathrm{step}}(t_{\mathrm{f}})>0$.
The normalized history and diffusive balance residuals are
\begin{subequations}
\label{eq:step_energy_residuals}
\begin{align}
\mathcal{R}_{j}^{\mathrm{hist}}(t_{\mathrm{f}})
&=
\frac{
\left|
\mathcal{W}_{j}^{\mathrm{step}}(t_{\mathrm{f}})
-
\mathcal{U}_{\alpha_j}^{\mathrm{hist}}(t_{\mathrm{f}})
-
\displaystyle
\int_{0}^{t_{\mathrm{f}}}
\mathcal{Q}_{\alpha_j}^{\mathrm{hist}}(t)\,dt
\right|}
{
\mathcal{W}_{j}^{\mathrm{step}}(t_{\mathrm{f}})},
\label{eq:step_history_energy_residual}
\\
\mathcal{R}_{j}^{\mathrm{diff}}(t_{\mathrm{f}})
&=
\frac{
\left|
\mathcal{W}_{j}^{\mathrm{step}}(t_{\mathrm{f}})
-
\mathcal{U}_{\alpha_j}^{\mathrm{diff}}(t_{\mathrm{f}})
-
\displaystyle
\int_{0}^{t_{\mathrm{f}}}
\mathcal{Q}_{\alpha_j}^{\mathrm{diff}}(t)\,dt
\right|}
{
\mathcal{W}_{j}^{\mathrm{step}}(t_{\mathrm{f}})}.
\label{eq:step_diffusive_energy_residual}
\end{align}
\end{subequations}
The continuum identities in
Secs.~\ref{sec:energy_decomposition} and
\ref{sec:diffusive_representation}
give
\begin{equation}
\mathcal{R}_{j}^{\mathrm{hist}}(t_{\mathrm{f}})
=
\mathcal{R}_{j}^{\mathrm{diff}}(t_{\mathrm{f}})
= 0.
\label{eq:exact_step_energy_residuals}
\end{equation}
Their difference reproduces the storage-gap identity.
\begin{equation}
\begin{aligned}
\mathcal{U}_{\alpha_j}^{\mathrm{hist}}(t_{\mathrm{f}})
-
\mathcal{U}_{\alpha_j}^{\mathrm{diff}}(t_{\mathrm{f}})
&=
\int_{0}^{t_{\mathrm{f}}}
\left[
\mathcal{Q}_{\alpha_j}^{\mathrm{diff}}(t)
-
\mathcal{Q}_{\alpha_j}^{\mathrm{hist}}(t)
\right]dt
\\
&\geq 0.
\end{aligned}
\label{eq:step_storage_gap_check}
\end{equation}
Equations~\eqref{eq:step_energy_residuals}--\eqref{eq:step_storage_gap_check} are analytic consistency identities, not an independent numerical validation. The parameter study evaluates the exact frequency and step responses for stated values of $\alpha$, $\tau$, $\chi$, $\epsilon_{\infty}$, and $\sigma$. It also checks the Mittag--Leffler quadrature against independent values. The results quantify fractional-order broadening, attenuation redistribution, the initial layer, and algebraic settling. They do not establish convergence of a time-stepping solution for the coupled Maxwell system. Such a study would need to specify the temporal mesh, singular quadrature, spectral nodes, and positive weights. It would also need to demonstrate convergence of both residuals, preserve $0\leq\mathcal{U}_{\alpha_j}^{\mathrm{diff}}\leq\mathcal{U}_{\alpha_j}^{\mathrm{hist}}$, and resolve the $t^{\alpha_j-1}$ initial layer.
The causal stationary limit of the finite-time harmonic solution supplies the permittivity used in the absorption and Maxwell eigenproblem calculations. The stated frequency limits and exact step identities agree with the constitutive and energy formulations. These results connect transient dielectric memory with stationary dispersion and attenuation within the finite-terminal Caputo model.

\section{Conclusion}
\label{sec:conclusion}

This study developed an energy formulation for a linear anisotropic Maxwell--Cole--Cole medium with multiple fractional polarization channels, passive conductivity, and prescribed external currents. Each channel obeys a finite-terminal Caputo constitutive law with relaxation time $\tau_j>0$, fractional order $0<\alpha_j<1$, and a uniformly positive-definite susceptibility tensor $\boldsymbol{\chi}_j$. Fractional memory acts only on the material polarization, while the macroscopic Maxwell equations retain their standard spatial form. The divergence constraints, local free-charge conservation, electromagnetic transmission conditions, and boundary power structure remain unchanged.

The polarization fields belong to a weighted temporal class that permits the fractional initial layer generated by quiescent Caputo data. For a nonzero initial electric field, the generic behavior is $\mathbf{P}_j(t)=\mathcal{O}(t^{\alpha_j})$ and $\partial_t\mathbf{P}_j(t)=\mathcal{O}(t^{\alpha_j-1})$. The polarization is continuous at initialization, but its time derivative may be unbounded. The energy theorem does not require the generally incompatible assumption $\mathbf{P}_j\in C^{2}([0,T])$. If classical regularity is imposed, the constitutive equation requires $\mathbf{E}(0)=\mathbf{0}$. Under the stronger assumptions $\mathbf{P}_j\in C^{2}$ and $\mathbf{E}\in C^{1}$, it also requires $\partial_t\mathbf{P}_j(0)=\mathbf{0}$ and $\partial_t\mathbf{E}(0)=\mathbf{0}$.

The Caputo convolution separates the fractional polarization power into the derivative of a non-negative history-storage functional and a non-negative constitutive dissipation rate. The construction uses squared polarization increments and does not invoke a generalized fractional product rule. The weakly singular integrals are first evaluated on truncated time intervals and then extended to the endpoints using the weighted temporal bounds. The history storage is absolutely continuous and vanishes at initialization. The instantaneous fractional dissipation may scale as $t^{\alpha_j-1}$ and diverge as $t\to0^{+}$, but it remains locally integrable. Its accumulated value over $[0,t]$ tends to zero. The weighted regularity gives order bounds for the initial storage and dissipation. Explicit leading coefficients require a uniform weighted small-time expansion of the Volterra solution.

Combining the material decomposition with the Poynting theorem gives a global field-plus-memory energy identity. The storage contains the electric and magnetic field energies, the instantaneous polarization-amplitude energy, and the fractional history energy. The dissipation contains conductive and fractional constitutive losses. The supply rate includes the inward boundary Poynting flux and the work supplied by the prescribed current. The integrated identity then gives a global passivity inequality. This integrated form is the primary theorem for solutions with a fractional initial layer. Its differential form holds only almost everywhere in time.

For a closed system without an external current, the total energy is non-increasing, and its decrease equals the accumulated passive dissipation. The same estimate gives conditional uniqueness under PEC, periodic, or another homogeneous passive boundary condition for which the difference of two solutions has zero boundary supply. The compared solutions must have identical material tensors, external sources, divergence constraints, initial electromagnetic fields, and zero polarization prehistory. The argument establishes uniqueness within the stated weak solution class if solutions exist. It is not an existence theorem.

The limiting cases require stated convergence hypotheses. For a fixed differentiable polarization history, the limit $\alpha_j\to1^{-}$ removes the fractional history storage and recovers the local Debye dissipation and anisotropic Debye constitutive law. For a family of coupled Maxwell solutions, this conclusion also requires convergence and uniform bounds away from the initialization layer. At fixed $0<\alpha_j<1$, the limit $\tau_j\to0^{+}$ gives the instantaneous polarization $\epsilon_0\boldsymbol{\chi}_j\mathbf{E}$ at each fixed positive time under suitable uniform history bounds. The convergence is generally nonuniform at $t=0$. Quiescent polarization with a nonzero initial electric field produces a shrinking relaxation layer whose accumulated dissipation need not vanish. A separate singular-limit analysis would be needed to determine whether this loss converges to a measure concentrated at initialization. A global instantaneous limit without an initial defect requires well-prepared polarization data or, within the zero-prehistory model, compatible zero initial electric data. The vanishing-susceptibility limit removes each channel continuously. Removing all polarization channels and setting $\boldsymbol{\epsilon}_{\infty}=\mathbf{I}$ and $\boldsymbol{\sigma}=\mathbf{0}$ recovers the vacuum Maxwell energy law.

The Caputo kernel also has a representation as a positive continuum of exponential relaxation modes. This standard spectral form produces auxiliary diffusive variables and a second non-negative storage and dissipation pair for the same fractional power. An exact comparison shows that the diffusive energy does not exceed the history energy. Their difference is a non-negative weighted variance of the polarization increments around the diffusive states. Equality holds only when the polarization history vanishes on the observation interval. The storage gap equals the difference between the accumulated diffusive and history dissipations, but the two instantaneous dissipation rates have no general pointwise ordering.

The two storage functionals are compatible but generally distinct. The variance gap measures the quadratic information lost when exponentially weighted moments represent the full polarization trajectory. It is not an additional physical energy reservoir. The ordering applies only to the two constructions derived here. It does not identify the diffusive energy as a minimum free energy, the history energy as a maximum free energy, or either functional as thermodynamically unique. This restriction is consistent with the nonuniqueness of storage and free-energy functionals in hereditary materials \cite{Fabrizio:2014}.

A finite positive quadrature of the relaxation spectrum preserves the non-negativity of the discrete diffusive energy and dissipation. It also preserves the passive sign structure of the continuum constitutive law. Positive nodes and weights alone do not establish approximation accuracy, kernel convergence, temporal-discretization convergence, or convergence of the coupled Maxwell system. These questions require separate numerical analysis.

The finite-time response to an abruptly switched harmonic field was obtained from the original Caputo initial-value problem. This calculation separates the fractional initialization transient from the stationary material response and does not replace the finite-terminal derivative by a separate infinite-prehistory constitutive operator. The Mittag--Leffler transient decays algebraically, and its causal long-time limit gives the anisotropic multi-term Cole--Cole permittivity with branch $(-i\omega\tau_j)^{\alpha_j}$. The Hermitian imaginary part of the effective permittivity is positive semidefinite, so the cycle-averaged material absorption is non-negative under the same susceptibility and conductivity assumptions that yield time-domain passivity.

The stationary source-free Maxwell eigenproblem was introduced only after the constitutive long-time limit and only for a spatially homogeneous medium. It determines the polarization eigenvectors and complex wave numbers of the anisotropic material. In the isotropic limit, it gives the phase constant, attenuation constant, and penetration depth. The relaxed tensor $\boldsymbol{\epsilon}_{s}$ is the static polarization permittivity after conduction has been separated. When $\boldsymbol{\sigma}\neq\mathbf{0}$, it is not the zero-frequency limit of the full effective permittivity because the conductive contribution diverges as $\omega^{-1}$. The Debye, instantaneous, and high-frequency limits agree with the corresponding time-domain constitutive limits and the standard Cole--Cole response \cite{ColeCole1941,GarrappaMainardiMaione2016}. The attenuation power laws apply only under weak loss, negligible conductivity, a slowly varying positive background, and nonoverlapping relaxation bands. They are not general anisotropic multichannel asymptotics.

The parameter study shows the role of $\alpha$. The loss maximum remains at $\omega\tau=1$, but its height is $\tfrac12\tan(\pi\alpha/4)$. Its half-maximum bandwidth grows from $1.1439$ decades in the Debye limit to $3.6756$ decades at $\alpha=0.4$. The exact attenuation curves cross because lowering $\alpha$ increases loss far from the characteristic frequency but decreases it near the peak. Under step excitation, a smaller $\alpha$ produces a stronger integrable initial power layer and a slower algebraic relaxation tail. The fractional order therefore controls the distribution of loss across time and frequency rather than acting as a uniform loss factor. The accompanying notebook specifies the illustrative parameters and reproduces the four single-panel figures.

The exact electric-step solution checks the two storage decompositions. It reproduces the fractional initial layer, gives non-negative constitutive power, and satisfies both integrated energy identities exactly. The corresponding residuals can test future history and spectral discretizations. Their exact vanishing follows from the continuous theory. The numerical component evaluates the exact constitutive responses and quantifies the effect of $\alpha$. It does not validate a fully discrete Maxwell algorithm.

The analysis applies to finite-terminal Caputo memory with zero prehistory, linear polarization, time-independent passive material tensors, spatially local constitutive behavior, and nonmagnetic media. The tensors may be discontinuous across Lipschitz material interfaces when Maxwell's equations are interpreted weakly and the standard transmission conditions are imposed. The formulation excludes nonlinear response, magnetic dispersion, spatial nonlocality, time-dependent or aging material parameters, initialized nonzero memory, thermal evolution, and heat transport. The non-negative constitutive loss is an electromagnetic energy-dissipation rate, not an entropy-production rate. A thermodynamic extension would require a coupled thermal balance, a positive temperature field, heat transport, and a Clausius--Duhem inequality. Within these assumptions, the energy balances connect fractional initial layers, history-dependent storage, electromagnetic passivity, diffusive relaxation states, stationary Cole--Cole dispersion, harmonic absorption, and attenuation.

\begin{acknowledgments}
The author acknowledges institutional support from the University of Santo Tomas.
\end{acknowledgments}

\section*{Data availability statement}
No external dataset was used. The numerical curves are generated directly from the manuscript equations. The accompanying \texttt{fractional\_order\_study\_R2.ipynb} notebook contains the parameters, quadrature, verification values, and plotting code.

\section*{Conflict of interest}
The author declares no conflict of interest.

\bibliography{ref}

\end{document}